\documentclass{aa}
\usepackage{txfonts}

\usepackage{graphicx}
\usepackage{natbib}

\begin{document}

\title{Optical and near-infrared recombination lines \\ 
       of oxygen ions from Cassiopeia~A knots%
\thanks{Tables 5 to 23 are only available in electronic form
at the CDS via anonymous {\tt ftp} to {\tt cdsarc.u-strasbg.fr}
({\tt 130.79.128.5})
or via {\tt http://cdsweb.u-strasbg.fr/cgi-bin/qcat?J/A+A/}}%
}

\titlerunning{Optical and NIR recombination lines
              from Cas A knots}

\author{D.\ Docenko\inst{1,2} \and R.A.\ Sunyaev\inst{1,3}
       }
\institute{Max Planck Institute for Astrophysics,
           Karl-Schwarzschild-Str.\ 1,
           85741 Garching, Germany
           \and
           Institute of Astronomy, University of Latvia,
           Rai\c{n}a bulv\={a}ris 19,
           Riga LV-1586, Latvia
           \and
           Space Research Institute, Russian Academy of Sciences, 
           Profsoyuznaya 84/32,
           117997 Moscow, Russia
           }




\abstract
{Fast-moving knots (FMK) in the Galactic supernova remnant Cassiopeia~A
consist mainly of metals and 
allow to study element production in supernovae and shock physics
in great detail.
}
{We work out theoretically and suggest to observe previously unexplored
 class of spectral lines -- metal recombination lines in optical and 
 near-infrared bands -- emitted by the cold ionized and 
 cooling plasma in the fast-moving knots.
}
{By tracing ion radiative and dielectronic recombination,
collisional $l$-redistribution and radiative cascade processes,
we compute resulting oxygen, silicon and sulphur 
recombination line emissivities. 
It allows us to determine the oxygen recombination line fluxes,
based on the fast-moving knot model of Sutherland and Dopita (1995b),
that predicts existence of highly-ionized ions from moderate to very low
plasma temperatures.
}
{The calculations predict oxygen ion recombination line fluxes
detectable on modern optical telescopes in the wavelength range
from 0.5 to 3~$\mu$m. 
Line ratios to collisionally-excited lines will
allow to probe in detail the process of rapid cloud cooling
after passage of a shock front,
to test high abundances of \element[4+]{O}, \element[5+]{O} 
and  \element[6+]{O} ions
at low temperatures and measure them,
to test existing theoretical models of a FMK 
and to build more precise ones.
}
{}


\maketitle

\section{Introduction}
\label{SecIntro}

The brightest source on the radio sky 
-- a supernova remnant Cassiopeia~A --
is so close and so young that the best instruments are able to observe
exceptionally fine details in its rich structure
in different spectral bands.
One of the most interesting phenomena are 
dense ejecta blobs of the supernova explosion, observed as
numerous bright optical ``fast-moving knots'' (FMKs).
These knots radiate mostly in [\ion{O}{iii}] doublet near 5000~\AA\
and other forbidden lines of oxygen, sulphur, silicon and argon.

Early observations using 200-inch telescope
on Mount Palomar~\citep{BaaMin54,MinkAll54,Mink57}
demonstrated that the emitting knot plasma has temperatures
about $2\times10^4$~K and electron densities about $10^4$~cm$^{-3}$.
Very unusual spectra of the fast-moving knots
have lead \citet{ShklovskyBook} to suggestion, later
confirmed by detailed measurements and analysis
\citep{Peimbert71,CheKirCasA,CheKirAbund},
that they have extremely high abundances of some of heavy elements.
Some of the knots consist of up to 90\% of oxygen, others contain
predominantly heavier elements, such as silicon, sulphur, argon
and calcium~\citep{CheKirAbund,CasAFesen96}.

It was also discovered that a typical lifetime of a FMK is of the
order of 10-30 years. Existing knots disappear, but other bright knots
appear on the maps of Cassiopeia A (hereafter Cas~A).

Detailed Chandra observations~\citep{ChandraCasA2000,CasACha1Ms}
revealed presence of similar knots in X-rays, being very bright
in spectral lines of hydrogen- and helium-like ions of silicon and
sulphur and showing traces of heavily absorbed oxygen X-ray lines. 

Similar bright optical knots are observed in other oxygen-rich
supernova remnants (e.g., Puppis~A, N132D, etc., \citet{SD95I}),
but the case of Cas~A is the most promising for further investigations.

The FMK optical spectra were first treated as
the shock wave emission by \citet{CheKirCasA,CheKirAbund}.
Subsequently it was understood that used shock models,
that assumed solar abundances,
are inappropriate for the case of the FMK plasma,
as its high metal abundances result in a different shock structure.

There have been published several theoretical models
describing the shock emission in pure oxygen
plasmas \citep{Itoh81a,Itoh81b,BorkowskiO} and oxygen-dominated plasmas
\citep[hereafter SD95]{SD95}.
In these models, the reverse shock of the supernova remnant
encounters a dense knot and decelerates entering its dense medium
\citep{ZeldRaiz66,McKeeCowie75}
to velocities of {several hundred km/s}.
Shortly after the shock wave enters the cloud,
the material ahead of the shock front is ionized
by the radiation from the heated gas after the shock front.
{All the theoretical models assume a shock wave propagation
with a constant velocity through a constant density cloud.}

The optical emission is formed in two relatively thin layers:
one immediately following
the shock front, where plasma rapidly cools from 
X-ray emitting temperatures of about $5\times10^6$~K
down to below $10^3$~K;
another at the photoionization front before the shock wave.
The relatively cold layers having temperatures below $10^5$~K
contribute most strongly to the line emission, as approximate
pressure equilibrium results
in much higher densities and emission measures of these regions.

Important detail is that \emph{the recombination time is much longer
than the plasma cooling time}. Therefore recombining and cooling plasma
simultaneously contains ions in vastly different ionization stages
at all temperatures down to several hundred Kelvins
{-- the plasma is in a non-equilibrium ionization state}.
It would be very important to find a way to confirm experimentally
this prediction of the computational models.

It is interesting to note that predictions of various
theoretical models differ in many ways.
For example, as we show in an accompanying article
(Docenko \& Sunyaev, to be submitted),
the model predictions of the fine-structure far-infrared line intensities
differ by several orders of magnitude.
We also show there that 
the models of SD95 and of \citet{BorkowskiO} are the best of available
ones in reproducing the far-infrared line emission
of the fast-moving knots, although each of them is only precise up to
a factor of several.

Nevertheless, these two theoretical models still have a lot of differences
between them.
For example, in the \citet{BorkowskiO} models the \element[4+]{O}
and higher-ionized species are essentially absent,
whereas in the SD95 model the \element[6+]{O} is the dominant ion
after the shock at temperatures above about 5000~K.
Unfortunately, \element[4+]{O}, \element[5+]{O} and \element[6+]{O}
ions have no collisionally-excited lines in visible and infrared ranges,
and their ultraviolet and soft X-ray lines
are almost or completely undetectable due to high interstellar absorption
on the way from the Cas~A.

From large differences between the theoretical model predictions
it is clear that they have lack of observational constraints
and more diagnostic information in form of various line ratios
is needed to pin down the true structure of fast-moving knots.
Part of such information can be obtained from the far-infrared lines.

\bigskip
In this paper we argue that still more information
may be obtained from metal recombination lines (RL) 
in optical and near-infrared spectral ranges.
These lines arise in the transitions between highly-excited levels 
(principal quantum number $n\approx5-10$) populated
by the processes of dielectronic and radiative recombination.
The RLs are emitted by \emph{all} ionic species,
including, for example, \element[4+]{O} and \element[5+]{O} that
have no other lines detectable from the Cas~A.

{Unfortunately, the collisional excitation process that is much more
efficient at exciting lines corresponding to transitions between lowest
excited states, is negligibly weak at excitation of high-$n$ levels.
Therefore, emissivities of the RLs are normally
several orders of magnitude weaker
than ones of the collisionally-excited lines, but, as we show in this
article, intensities of the recombination lines from the FMKs are reaching
detectable levels.}

For oxygen-dominated knots, using plasma parameters
(electron temperature, density and emission measure of the gas
 at different distances from the shock wave)
from the SD95 model with the 200~km/s shock speed
we have computed the spectrum in the optical and near-infrared
bands accessible by the modern ground-based telescopes.
{The cloud shock speed in the Cas~A FMKs of about 200~km/s
is determined by SD95 from the comparison of observed optical spectra
with their model predictions.}
Although similar analysis based on the \citet{BorkowskiO} model
would also be valuable, the original article does not give enough
information to allow us to compute the RL fluxes 
{(e.g., the ionization state distributions as functions
of temperature cannot be resolved from the plots in the regions
significantly contributing to the recombination line emission)}.

The lines under consideration are predicted to be $300-500$ times
weaker than e.g.\ the [\ion{O}{iii}] line at 5007~\AA.
However, their fluxes are well above the sensitivity of the best
present-day telescopes.
These fluxes are only 2--5 times weaker than the detection limits
of observations done in 1970's and later on the 200-inch and smaller
telescopes~\citep{Peimbert71,CheKirAbund,CasAFesen96}.
Observations of these lines do not demand high spectral resolution
due to velocity spreads and turbulence in the shocked plasma,
as well as due to intrinsic line splitting.

The recombination lines discussed below are 
radiated by all coexisting ionized oxygen species, from 
\element[+]{O} to \element[6+]{O}.
Study of these lines will permit
  to observe in details the process of the non-equilibrium cooling
and recombination of the pre- and post-shock
plasmas overabundant with oxygen or sulphur and silicon.

The discussed spectral lines allow to observe the coldest ionized pre-shock
and post-shock regions in the FMKs, that are not possible to detect
using the ground-based observations by any other means.
The fine-structure far-infrared emission lines 
are also able to give such information, but
demand an observatory to be located outside the atmosphere
and cannot reach angular resolution sufficient
to distinguish individual knots.

Observations of recombination lines of elements other than oxygen will
allow to improve our knowledge on many yet unknown physical parameters
characterizing the Cas~A ejecta.

The metal recombination lines may be a promising tool also
for studies of other oxygen-rich supernova remnants 
{like Puppis~A, N132D, G292+1.8, etc}.

\bigskip

The paper structure is following. In the second Section we 
describe our method of the recombination line flux computation.
In the Section~\ref{SecResults} we apply the method to predict
the optical and near-infrared line fluxes
from the optical knots in Cassiopeia~A
and compare them with existing observational constraints.
Description of the recombination line substructure
allowing to identify the parent ion
is given in the Section~\ref{SecLineStr}.
In the Section~\ref{SecModResults} we discuss the ways to
get information about physical parameters of the emitting region
from the line ratios.
Finally, in Section~\ref{SecConclusions}
we conclude the article.

\section{Computation of recombination line fluxes}
\label{SecORL}

In the process of radiative recombination, especially at low
temperatures 
(when $kT$ is at least several times below the ionization potential),
significant part of electrons recombine onto excited states.
Dielectronic recombination for majority of heavy ions populate
excited states even more efficiently than the radiative recombination.

In low-density plasmas such recombination onto excited states,
characterized by quantum numbers $(nl)$,
is followed by the electron radiative cascade to the ground state
and emission of multiple photons in the course of the cascade.
In case of recombining highly-charged ions, these photons
are emitted in microwave, infrared, optical and ultraviolet
spectral bands.

In this paper we describe the lines produced as a result of such radiative
cascade, when both collisional and induced transitions of the type
$(nl)\to(n'l')$ are unimportant. The only non-radiative process
influencing electron cascade in our model
are the collisional $(nl)\to(nl')$ transitions,
{as they have much higher cross sections and significantly
affect the level populations at relatively low $n$.
They also are the main reason of changes of the recombination line
emissivities with electron density.}

We show in Appendix~\ref{AppA} that the $n$-changing transitions
may be neglected for computations of optical recombination line
fluxes, while still retaining reasonable accuracy of the results
{(better than 20\%)}.

\subsection{Elementary processes}
\label{SecProc}

In the current work we account for the following processes:
\begin{itemize}

\item Radiative recombination (RR). Its level-specific rates
$q_{\rm RR}(nl;T_{\rm e})$ were computed 
{as described in Appendix~\ref{SecRR}.}

\item Dielectronic recombination (DR). Its level-specific rates
$\sum_\gamma q_{\rm DR}(\gamma,nl;T_{\rm e})$ were computed as
described in Appendix~\ref{SecDR}.

\item Radiative transitions. Their rates $A_{nl,n'l'}$ were computed
  using hydrogenic formulae with radial integral
  expressions from \citet{Gordon29}.
  This may be not accurate for non-hydrogenic ions at $l<3$, 
  but electrons recombined to those low-$l$ states mostly transit to
  $n'\approx l$ and do not contribute to the optical RL emission%
\footnote{{
After all computations have been performed, the authors were
informed about a more elegant and modern way of hydrogenic radial
integral computation using associated Laguerre polynomials
(\citet{MalikRnl}; note that there is a typographical error
in one of their equations, which is rectified in the appendix
of \citet{Kevin07}).}}.

\item Collisional $l$-redistribution.
  Its rates $C_{nl,nl'}$ were computed using sudden
  collision approximation expressions from~\citet{PengellySeaton}
  and \citet{Summersbnl}, in the region of their applicability.
  Outside it (where maximum cut-off parameter
  $r_{\rm max}$ becomes less than the minimum one $r_{\rm min}$), 
  we approximated the~\citet{VrinStark} classical expressions introducing
  cut-off at large impact parameters equal to $r_{\rm max}$.
  Details of this approximation are of minor importance, since
  the cross sections are proportional as the $r_{\rm max}^2$,
  being therefore relatively small.
  Energies of $(nl)$-levels and their differences determining
  the collisional rates in many cases, 
  were computed as described in Appendix~\ref{SecQuaDef}.
\end{itemize}

Usage of hydrogenic approximation does not allow us to reliably 
compute emissivities and wavelengths of the lines corresponding
to transitions involving levels $n<4$.
Therefore we do not provide the results concerning these levels.

Details of the atomic physics and approximations utilized
for computations of the line emissivities are given in
Appendix~\ref{AppB}.

\subsection{Cascade and $l$-redistribution equations}

The equation describing radiative cascade of a highly-excited electron,
accounting for the collisional $l$-redistribution, is 
\begin{equation}
\label{EqPopnl}
\begin{array}{l}
 N_{nl} \left(  \sum_{n'<n} \sum_{l'=l\pm 1} A_{nl,n'l'}  
              + \sum_{l' \ne l} C_{nl,nl'}\right) = {}\\[5pt]
\qquad            = n_{\rm e} n_+ \left(q_{\rm RR}(nl)
                         + \sum_\gamma q_{\rm DR}(\gamma, nl)\right) \\[5pt]
\qquad\qquad     + \sum_{n'>n}\sum_{l'=l\pm1} N_{n'l'} A_{n'l',nl}
                 + \sum_{l'\ne l} N_{nl'} C_{nl',nl},
\end{array}
\end{equation}
where $n_{\rm e}$, $n_+$ and $N_{nl}$ are the number densities of electrons,
recombining ions and recombined ions with electron on the level $(nl)$.

Analytic solution of this coupled system of equations
involves inversion of large matrices
of sizes of up to several tens of thousands
and is not feasible.

We solved the problem the following way.
Starting from some maximum $n=n_{\rm max}$, relevant for the problem
(defined in the Appendix~\ref{AppA}),
we neglected its cascade population.
Knowing the level-specific recombination and $l$-redistribution rates,
we computed resulting populations of levels $(nl)$ by numerical solution
of the system of linear equations~(\ref{EqPopnl}) for $n=n_{\rm max}$.

Using level-specific radiative rates, we could compute 
$l$-resolved cascade population of the levels $(nl)$
for $n=n_{\rm max}-1$.
Having computed total population rates, we solved the
system~(\ref{EqPopnl}) for this $n$.
In such a way, moving downwards in $n$, populations of all levels
were computed.

\subsection{Recombination line emissivities}
\label{SecLineEps}

Having obtained level populations $N_{nl}$,
we have computed recombination line emissivities $\varepsilon(nl,n'l')$,
cm$^3$/s, defined as
\begin{equation}
\label{EqEmisDef}
 \varepsilon(nl,n'l') = \frac{{\rm d}{\mathcal N}(nl\to n'l')}{{\rm d}V{\rm d}t}  
                       \frac{1}{n_{\rm e} n_+}
                      = \frac{N_{nl}}{n_{\rm e} n_+} A_{nl,n'l'},
\end{equation}
where d${\mathcal N}(nl\to n'l')/{\rm d}V{\rm d}t$
is the number of transitions from level $(nl)$ to $(n'l')$
per second in the unit volume.

\begin{figure}
\begin{center}
\centerline{
    \rotatebox{270}{
        \includegraphics[height=0.95\linewidth]{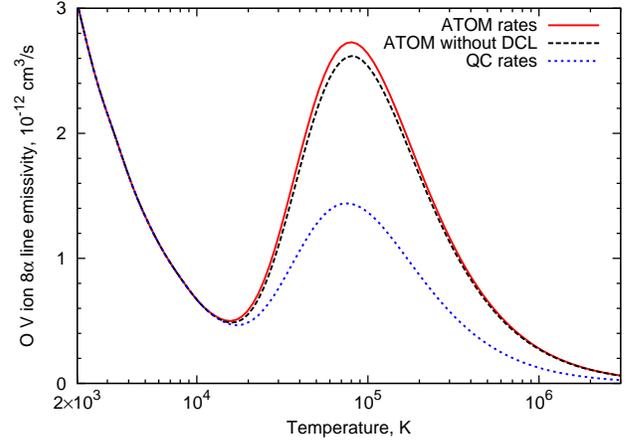}
                   }
           }
\caption{Temperature dependence of the \ion{O}{v} 8$\alpha$
         recombination line emissivity in the low-density
         limit. Lower curve correspond to DR rate computations
         using quasi-classical autoionization rates, 
         upper curve -- to the ATOM rates 
         (for details see Appendix~\ref{SecDR}),
         middle curve does not account for the process of
         line emission without recombination (DCL),
         described in Appendix~\ref{SecDCL}.
         At low temperatures the emissivity increase
         is determined by the radiative recombination, but
         the peak around $10^5$~K is arising
         due to the dielectronic recombination.
         }
   \label{FigEmiss8a}
  \end{center}
\end{figure}

As an example, predicted \ion{O}{v} 8$\alpha$ line%
\footnote{%
As it is usual, by $n\alpha$ we denote spectral line formed by electronic
transition from level $n+1$ to level $n$,
by $n\beta$ -- from $n+2$ to $n$, etc.
}
low-density emissivity with $l$- and $l'$-components summed up 
is shown on Figure~\ref{FigEmiss8a}
as a function of temperature for different approximations.

Emissivities of recombining \element[5+]{O} 
optical and near-infrared recombination lines
are given in Table~\ref{TabOVIeT}
as functions of temperature in the low-density limit.

Approximate wavelengths%
\footnote{Here and everywhere below we present the vacuum wavelengths.}
of the brightest lines 
of all ions as functions of ionization stage are given in
Table~\ref{TabRLlam}.
They can be easily estimated using
hydrogenic expressions for any recombination line arising
in transition $n\to n'$ between levels with $n'>3$, $l>2$ as
\begin{equation}
 \lambda(n\to n') \approx \frac{0.091127}{Z^2}\,
    \left(\frac{1}{n'^2} - \frac{1}{n^2} \right)^{-1}\, \mu{\rm m},
\end{equation}
{where $Z$ is the recombining ion charge.}

\begin{table}
\caption{%
Wavelengths $\lambda$ and emissivities $\varepsilon$ of several
\ion{O}{v} optical and near-infrared recombination lines
in the low-density limit.
\label{TabOVIeT} 
}
\begin{center}
\begin{tabular}{rc|llll}
\hline\hline
 & &\multicolumn{4}{c}
{Emissivity $\varepsilon$, cm$^{3}$/s, for electron temperature $T_{\rm e}$}\\
Line & $\lambda$, $\mu$m & $1\cdot10^3$~K & $3\cdot10^3$~K & 
                           $1\cdot10^4$~K & $3\cdot10^4$~K \\
\hline
$ 5\alpha$ & 0.298  & $9.4\cdot10^{-12}$ & $4.1\cdot10^{-12}$ &
 $3.5\cdot10^{-12}$ & $5.9\cdot10^{-12}$\\
$ 6\alpha$ & 0.495  & $7.6\cdot10^{-12}$ & $3.2\cdot10^{-12}$ &
 $1.2\cdot10^{-12}$ & $3.0\cdot10^{-12}$\\
$ 7\alpha$ & 0.762  & $6.3\cdot10^{-12}$ & $2.5\cdot10^{-12}$ &
 $8.8\cdot10^{-13}$ & $1.9\cdot10^{-12}$\\
$ 8\alpha$ & 1.112  & $5.2\cdot10^{-12}$ & $2.0\cdot10^{-12}$ &
 $6.7\cdot10^{-13}$ & $1.3\cdot10^{-12}$\\
$ 9\alpha$ & 1.554  & $4.4\cdot10^{-12}$ & $1.7\cdot10^{-12}$ &
 $5.2\cdot10^{-13}$ & $8.8\cdot10^{-13}$\\
$10\alpha$ & 2.100  & $3.7\cdot10^{-12}$ & $1.4\cdot10^{-12}$ &
 $4.1\cdot10^{-13}$ & $6.0\cdot10^{-13}$\\
$11\alpha$ & 2.761  & $3.2\cdot10^{-12}$ & $1.1\cdot10^{-12}$ &
 $3.3\cdot10^{-13}$ & $4.0\cdot10^{-13}$\\
$12\alpha$ & 3.548  & $2.7\cdot10^{-12}$ & $9.3\cdot10^{-13}$ &
 $2.6\cdot10^{-13}$ & $2.6\cdot10^{-13}$\\
\hline
\end{tabular}
\end{center}
\end{table}

\begin{table}
\caption{%
Hydrogenic vacuum wavelengths, $\mu$m, of some optical and near-infrared
recombination $\alpha$-lines for several ionization stages,
denoted by the ion spectroscopic symbol after recombination.
\label{TabRLlam} 
}
\begin{center}
\begin{tabular}{r|cccccc}
\hline\hline
Line       & I & II & III & IV & V & VI \\
\hline
 $4\alpha$ & 4.0501 & 1.0125 & 0.4500 & 0.2531 & 0.1620 & 0.1125\\
 $5\alpha$ & 7.4558 & 1.8640 & 0.8284 & 0.4660 & 0.2982 & 0.2071\\
 $6\alpha$ & 12.365 & 3.0913 & 1.3739 & 0.7728 & 0.4946 & 0.3435\\
 $7\alpha$ & 19.052 & 4.7629 & 2.1168 & 1.1907 & 0.7621 & 0.5292\\
 $8\alpha$ & 27.788 & 6.9471 & 3.0876 & 1.7368 & 1.1115 & 0.7719\\
 $9\alpha$ & 38.849 & 9.7122 & 4.3165 & 2.4281 & 1.5540 & 1.0791\\
$10\alpha$ & 52.506 & 13.127 & 5.8340 & 3.2817 & 2.1003 & 1.4585\\
$11\alpha$ & 69.035 & 17.259 & 7.6705 & 4.3147 & 2.7614 & 1.9176\\
$12\alpha$ & 88.707 & 22.177 & 9.8563 & 5.5442 & 3.5483 & 2.4641\\
\hline
\end{tabular}
\end{center}
\end{table}

\subsection{Resulting line fluxes}
\label{SecLineI}

Fluxes $I(nl,n'l')$, erg/cm$^2$/s, of the lines
were finally computed by integrating along the line of sight
\begin{equation}
\label{EqInt}
 I(nl,n'l') = h\nu\frac{S}{4\pi R^2}
     \int \varepsilon\left(nl,n'l'; T_{\rm e}(r)\right) 
        n_{\rm e}(r) n_+(r) {\rm d}r,
\end{equation}
where $h\nu$ is the photon energy, $R$ is the distance from 
the observer to the emitting region and $S$ is the emitting region area.

The integral over distance in the plain-parallel approximation
can be easily transformed into integral over temperature by substitution
$$
 {\rm d}r = \frac{{\rm d}r}{{\rm d}t} \frac{{\rm d}t}{{\rm d}T} {\rm d}T
          = \upsilon_{\rm shock} \frac{3/2 (n_{\rm t}+n_{\rm e}) k_B}
                               {n_{\rm t} n_{\rm e} \Lambda_N} {\rm d}T,
$$ 
where $\upsilon_{\rm shock}$ is the shock front speed,
      $\Lambda_N$ is a cooling function and
      $n_{\rm t}$ is the total number density of all ions in plasma.

The parameters of this equation -- 
cooling function, electron and ion densities as functions of temperature
-- were taken from the SD95 200 km/s shock model,
described in more details in Section~\ref{SecResults} below.

For purpose of qualitative analysis, we introduce below the
oxygen differential emission measure per logarithmic temperature
interval
$$
\frac{{\rm d}\,E_{\rm O}}{{\rm d}\, (\log T_{\rm e})} = 
  T_{\rm e} \, \frac{n_{\rm O} n_{\rm e} {\rm d}r}{{\rm d}T_{\rm e}},
$$
where $n_{\rm O}$ is the total oxygen ion number density.
It shows contribution of a given logarithmic temperature interval to the
total emission measure, showing where most of the line emission originates.
Using this notion, we can express the line flux from Eq.~(\ref{EqInt}) as
$$
 I(nl,n'l') =  h\nu\frac{S}{4\pi R^2}
     \int \varepsilon\left(nl,n'l'; T_{\rm e}\right) 
          \frac{{\rm d}\,E_{\rm O}}
               {{\rm d}\, \log T_{\rm e}}
          \;
          \frac{n_+}{n_{\rm O}}
           T_{\rm e} {\rm d}T_{\rm e}.
$$
Here the fraction ${n_+}/{n_{\rm O}}$ is abundance of a given ionic species
and is also dependent on temperature.

In our computation of the post-shock recombination line emission,
we artificially stop integrating expression~(\ref{EqInt})
when plasma temperature drops below 1000~K.
This results in some, possibly significant, underestimate
of the cooling region line fluxes arising from recombinations
of \element[2+]{O} ions.
Though, as we show below, the \ion{O}{ii} recombination lines
from the photoionized region are expected to be much brighter.

\emph{It should be remembered} that, e.g., \ion{O}{v} recombination lines
arise in the radiative cascade in the \element[4+]{O} ion 
triggered by the recombination of \element[5+]{O} ion,
so the line fluxes are proportional to the ionic abundance
of \element[5+]{O}, not \element[4+]{O}.

\section{Astrophysical application: FMKs in Cassiopeia~A}
\label{SecResults}

\begin{figure}
\begin{center}
\centerline{
        \includegraphics[width=0.7\linewidth]{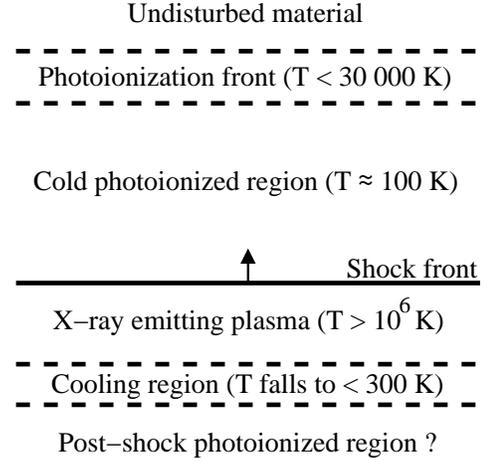}
           }
\caption{The plain-parallel SD95 model schematic structure.
         The drawing is not to scale. Arrow shows the shock wave direction.
         Optical lines are emitted from the photoionization front
         and the post-shock cooling region at temperatures of several
         tens of thousands Kelvin.
         Recombination lines arise in the cold photoionized region
         and the post-shock cooling region.
        }
   \label{FigSD95fig}
  \end{center}
\end{figure}

The SD95 model describes fast-moving knot emission
as arising in the interaction of the oxygen-dominated dense cloud
with the external shock wave, entering the cloud and propagating
through it (see Figure~\ref{FigSD95fig}).
The heated region just after the shock front produces
high flux of the ionizing radiation that results in
appearance of two photoionized regions: before and after the shock
front.

The plasma after the shock front passage
is rapidly cooling and, at temperatures of $(1-5)\times 10^4$~K,
emitting the lines observable in the visible and near-infrared spectra
\citep{CheKirCasA,CheKirAbund,CasAFesen96,CasANIR01}.
Thickness of this emitting layer is extremely small
(about $10^{12}-10^{13}$~cm),
but due to the high density of the cooling material 
at $T<10^5$~K it is able to produce bright emission lines
of highly-ionized atomic species, such as [\ion{O}{iii}].
This region also gives rise to the recombination line (RL) emission.

The SD95 model does not include
the photoionized region after the shock front.
Until now, even the presence of this region is disputed \citep{Itoh86}
because, if present, it would produce too bright optical lines
of neutral oxygen.

Between the photoionization front and the shock wave the plasma is predicted
to be extremely cold ($T_{\rm e}<10^3$~K) and rather highly ionized 
(ionic abundances of \element[3+]{O} and \element[2+]{O} are about 60\% and 
30\% in the SD95 200~km/s shock model).
This cold ionized region, invisible in optical and X-ray bands,
should be actively recombining and emitting bright recombination lines.

Resulting RL emission computed based on the SD95 200~km/s shock model
is described separately in the following subsections
for each of these two regions: after and before the shock front.

In this Section we provide results on the $l$-summed
recombination line fluxes, whereas in the Section~\ref{SecLineStr}
we discuss in details the individual line spectral substructure,
that separates in some cases the recombination lines into several
components, thus diminishing the individual component fluxes
up to factors of 2--4.

\subsection{Cooling region after the shock front}

{It is useful to make order-of-magnitude estimates
of some timescales in the post-shock rapidly cooling plasma
at temperatures between several hundred and $3\times10^4$~K,
where the recombination line emission occurs (see below).
As derived from SD95 model, the plasma cooling time is roughly equal to
$$
\tau_{\rm cool} = \frac32 \frac{(n_{\rm t}+n_{\rm e}) kT_{\rm e}}
                               {n_{\rm t} n_{\rm e} \Lambda}
   \approx 2\times10^4 \;{\rm s}\; T_4^{1/2}\, n_{\rm t,5}^{-1},
$$
where $T_{\rm e}=10^4T_4$~K and $n_{\rm t}=10^5n_{\rm t,5}$~cm$^{-3}$.
It should be compared with the time needed to reach the collisional
ionization equilibrium (CIE), that is in our case
is approximately equal to the recombination timescale $\tau_{\rm rec}$.
For example, for ion \element[2+]{O} and  \element[6+]{O}
the recombination time is
$$
  \tau_{\rm rec}({\rm O}^{6+}) = 
     \left(n_{\rm e} q_{\rm RR, O^{6+}} (T_{\rm e})\right)^{-1}
  \approx 4.7\times10^5 \;{\rm s}\; T_4^{0.64}\, n_{\rm e,5}^{-1},
$$
$$
  \tau_{\rm rec}({\rm O}^{2+}) = 
     \left(n_{\rm e} q_{\rm RR, O^{2+}} (T_{\rm e})\right)^{-1}
  \approx 4.9\times10^6 \;{\rm s}\; T_4^{0.64}\, n_{\rm e,5}^{-1},
$$
where $n_{\rm e}=10^5n_{\rm e,5}$~cm$^{-3}$ and $q_{\rm RR, O^{2+}}$
is the total radiative recombination rate for this ion, taken from
\citet{RRC91}.
It is easy to see that the recombination timescale in this temperature
interval is always longer than the cooling time for any abundant ion.
Therefore the plasma is strongly overionized for its temperature,
that results in high emissivities of the recombination lines.}

\begin{figure}
\begin{center}
\centerline{
    \rotatebox{270}{
        \includegraphics[height=0.95\linewidth]{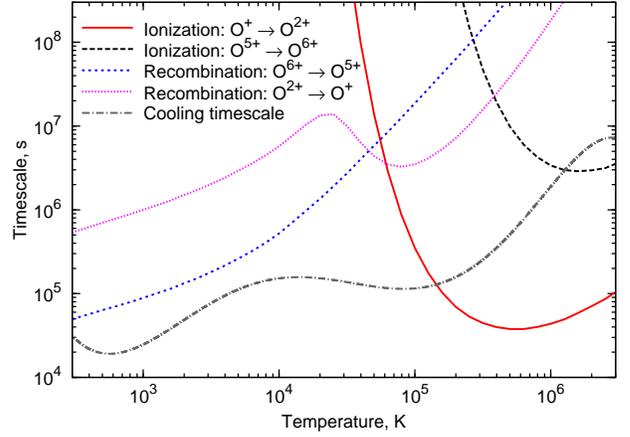}
                   }
           }
\caption{{Oxygen ion \element[6+]{O} and \element[2+]{O} 
        recombination and ionization timescales 
        as compared to the plasma cooling time
        using the electron density dependence on temperature
        and the cooling function from
        the SD95 model with 200~km/s cloud shock speed.}
        }
   \label{FigTaus}
  \end{center}
\end{figure}

{The quantitative comparison of these timescales in the model
is given on Figure~\ref{FigTaus}, where ionization and recombination
timescales for several ions are compared with the cooling time.
At high temperatures the cooling timescale is longer than the
ionization timescales and the material has enough time to 
converge to the CIE.
At $T_{\rm e}<10^6$~K the cooling is much faster than
the recombination down to temperatures of several hundred Kelvin
and the high ionization state is stays ``frozen'' until very low
temperatures are reached.}

{Let us now compare the thermalization timescales and mean free paths
to the quantities determined above.
Following \citet{Spitzer56}, the electron or ion self-collision timescale
is equal to 
$$
 \tau_{\rm c}= 114\;{\rm s}\;\frac{A^{1/2} T_4^{3/2}}
                                  {n_5 Z^4\ln\Lambda_{\rm C}},
$$
where $A$ is the atomic weight (equal to 1/1836 for electrons),
$Z$ is the particle charge, $n=10^5n_5$~cm$^{-3}$ 
is the particle concentration and 
$\Lambda_{\rm C}\approx3.92\times10^7\; n_{\rm e,5}^{-1/2} T_4^{3/2}$.
In our case of the highly-charged plasma having $Z\gg1$
the timescale of the ion-ion collisions is even shorter
than for the electron-electron collisions.}

{Nevertheless, the mean free path even for electrons $\lambda_{\rm e}$
stays significantly below any other characteristic scale,
including the typical temperature change
length scale of more than $10^{12}$~cm:
$$
\lambda_{\rm e}= \tau_{\rm c,e}\sqrt{3kT_{\rm e}/m_e}
 \approx 1\times10^7 \;{\rm cm}\; T_4^{2}\, n_{\rm e,5}^{-1}.
$$
}

\begin{figure}
\begin{center}
\centerline{
    \rotatebox{270}{
        \includegraphics[height=0.95\linewidth]{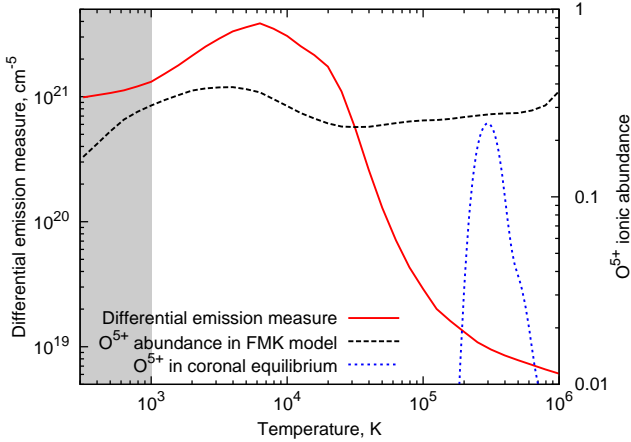}
                   }
           }
\caption{Differential oxygen emission measure
         per logarithmic temperature interval
         d$E_{\rm O}$/d$(\log T_{\rm e})$ in the post-shock cooling region
         of the SD95 FMK model.  
         Ionic abundances $n$(\element[5+]{O})$/n({\rm O})$ 
         of the \element[5+]{O} ion in the FMK model and in the collisional
         ionization equilibrium in low-density plasmas~\citep{Mazzotta}
         are also shown.
         Temperature range not taken into account in our calculations
         is shaded.
        }
   \label{FigEMSD95}
  \end{center}
\end{figure}

For illustration of the characteristic conditions after the passage
of the shock front, on Figure~\ref{FigEMSD95} we show
Li-like \element[5+]{O} ion density as function of plasma temperature
for this region%
\footnote{{
We made our own computations of the post-shock plasma recombination
and discovered that our oxygen ion distribution over ionization stages
is rather similar to the one presented on the lower panel of Figure~3 of SD95,
but only if the ion spectroscopic symbols on that figure
are increased by unity.
Therefore we assume that the Figure~3 of SD95 paper has a misprint
and the ion spectroscopic symbols should be read, e.g., 
``O~VII'' instead of ``O~VI'', ``O~VI'' instead of ``O~V'', etc.
}}.
The figure demonstrates incredibly high abundance of this ion
in relatively dense plasma with $T<3\times10^4$~K after the shock front,
much higher than in the CIE,
where this ion exists only in a narrow temperature range
$T_{\rm e}=(2-7)\times10^5$~K~\citep{Mazzotta}.

On the same figure we present also the oxygen emission measure 
distribution over temperature d$(EM)/$d(log$T_{\rm e}$) in the SD95 model.
This distribution together with the line emissivity dependence
on temperature (e.g., Fig.~\ref{FigEmiss8a})
allows one to calculate the relative contributions to 
the total line flux from different temperature intervals
(see Section~\ref{SecLineI} above).

\begin{figure}
\begin{center}
\centerline{
    \rotatebox{270}{
       \includegraphics[height=0.95\linewidth]{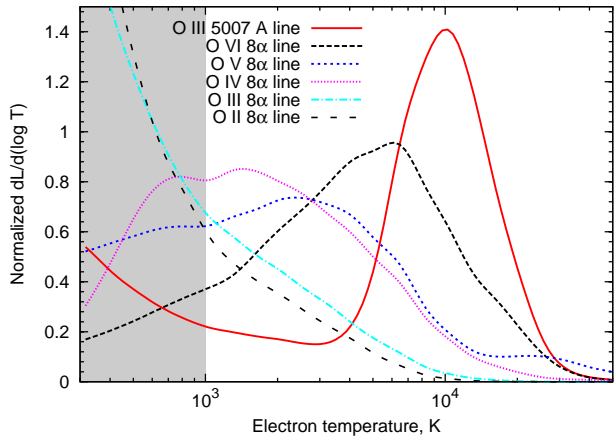}
                   }
           }
\caption{Differential luminosity distribution per logarithmic
         temperature interval according to the SD95
         model for several spectral lines,
         including prominent 5007~\AA\ [\ion{O}{iii}] line and predicted 
         \ion{O}{ii} -- \ion{O}{vi} 8$\alpha$ lines.
         Curves are normalized so that
         the area enclosed under each of them equals one.
         Temperature range not taken into account in our calculations
         is shaded.
         }
   \label{FigdLdT}
  \end{center}
\end{figure}

Such line flux distributions for
the prominent [\ion{O}{iii}] line at 5007~\AA\ and
8$\alpha$ lines of different oxygen ions are shown on Figure~\ref{FigdLdT}.
{Note that at low temperatures (below 4000~K), the [\ion{O}{iii}]
5007~\AA\ line emission is dominated by contribution from the recombinaiton
cascades, not the collisional excitation. The cascade contribution
to the post-shock intensity of this line reaches about 30\%.}

{Because of high \element[6+]{O} abundances
at both low and high temperatures,
the distribution of the \ion{O}{vi} 8$\alpha$ line emission
is much wider than that of the collisionally-excited 
optical [\ion{O}{iii}] line and has 90\% of its area within temperature range
  $5\times10^2-1.5\times10^4$~K.
Note that the dielectronic recombination is contributing
less than 20\% to the resulting line intensity for any of oxygen ions
in the discussed SD95 model.}

{
The \ion{O}{v} and \ion{O}{vi} lines are the brightest expected ones.
Lines of \ion{O}{iii} and \ion{O}{iv} from this region 
are predicted to be dimmer by a factor of about three, 
and the \ion{O}{ii} lines -- by a factor of about ten.
}

\begin{figure}
\begin{center}
\centerline{
    \rotatebox{270}{
       \includegraphics[height=0.95\linewidth]{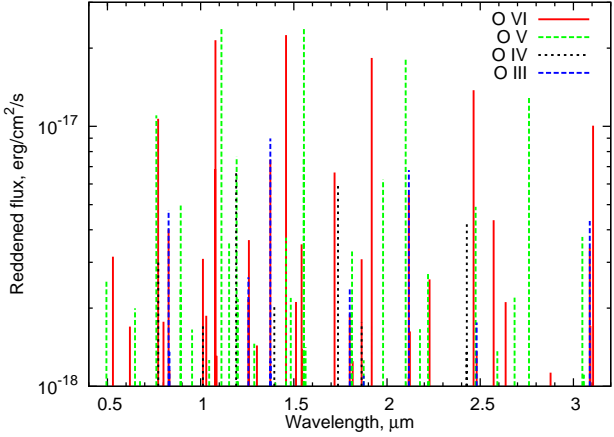}
                   }
           }
\caption{Reddened recombination line $l$-summed fluxes $F$
         from the post-shock cooling region.
         Details are as in note to Table~\protect\ref{TabFluxSW}.
         Here the lines are represented as ``infinitely thin'';
         their fine structure is discussed in Section~\protect\ref{SecLineStr}.
         }
   \label{FigSpeSD95}
  \end{center}
\end{figure}

On Figure~\ref{FigSpeSD95} we show the predicted oxygen recombination line
fluxes from the post-shock region of a model FMK.
The brightest recombination lines around 1~$\mu$m are about
300~times less intense than the reddened [\ion{O}{iii}] 5007~\AA\ line.
Expected flux values are given also in Table~\ref{TabFluxSW}.

\begin{table}
\caption{Brightest recombination line $l$-summed 
         emitted and reddened fluxes $I$ and $F$
         from the post-shock cooling region
         computed following the SD95 model.
         The optical [\ion{O}{iii}] 5007~\AA\ line flux also given.
         The last column contains the flux part in the strongest 
         unresolved line component $f_{\rm s}$.
         The wavelength $\lambda$ corresponds to the strongest line component.
\label{TabFluxSW}
}
\begin{center}
\begin{tabular}{lr|llll}
\hline\hline
Ion & Line & $\lambda$, $\mu$m &  $I$, erg/cm$^2$/s & $F$, erg/cm$^2$/s & $f_{\rm s}$\\
\hline
\ion{O}{vi} &  7$\alpha$ & 0.5293 & 3.43$\times10^{-16}$  &  3.15$\times10^{-18}$& 0.99  \\
\ion{O}{vi} &  8$\alpha$ & 0.7720 & 1.83$\times10^{-16}$  &  1.07$\times10^{-17}$& 0.99  \\
\ion{O}{vi} &  9$\alpha$ & 1.0792 & 1.03$\times10^{-16}$  &  2.14$\times10^{-17}$& 1.31$^*$\\
\ion{O}{vi} & 12$\beta $ & 1.3740 & 2.19$\times10^{-17}$  &  7.42$\times10^{-18}$& 0.99  \\
\ion{O}{vi} & 10$\alpha$ & 1.4586 & 6.08$\times10^{-17}$  &  2.24$\times10^{-17}$& 1.00  \\
\ion{O}{vi} & 13$\beta $ & 1.7189 & 1.49$\times10^{-17}$  &  6.65$\times10^{-18}$& 1.00  \\
\ion{O}{vi} & 11$\alpha$ & 1.9177 & 3.72$\times10^{-17}$  &  1.83$\times10^{-17}$& 1.00  \\
\ion{O}{vi} & 12$\alpha$ & 2.4642 & 2.35$\times10^{-17}$  &  1.38$\times10^{-17}$& 1.00  \\
\hline
\ion{O}{v}  &  6$\alpha$ & 0.4943 & 4.07$\times10^{-16}$  &  2.56$\times10^{-18}$& 0.62  \\
\ion{O}{v}  &  7$\alpha$ & 0.7618 & 1.99$\times10^{-16}$  &  1.10$\times10^{-17}$& 0.81  \\
\ion{O}{v}  &  8$\alpha$ & 1.1114 & 1.06$\times10^{-16}$  &  2.36$\times10^{-17}$& 0.89  \\
\ion{O}{v}  & 10$\beta $ & 1.1929 & 2.91$\times10^{-17}$  &  7.66$\times10^{-18}$& 0.92  \\
\ion{O}{v}  &  9$\alpha$ & 1.5538 & 5.92$\times10^{-17}$  &  2.36$\times10^{-17}$& 1.21$^*$\\
\ion{O}{v}  & 12$\beta $ & 1.9784 & 1.24$\times10^{-17}$  &  6.26$\times10^{-18}$& 0.96  \\
\ion{O}{v}  & 10$\alpha$ & 2.1002 & 3.47$\times10^{-17}$  &  1.83$\times10^{-17}$& 0.94  \\
\ion{O}{v}  & 11$\alpha$ & 2.7613 & 2.12$\times10^{-17}$  &  1.32$\times10^{-17}$& 0.97  \\
\hline
\ion{O}{iv} &  7$\alpha$ & 1.1902 & 2.52$\times10^{-17}$  &  6.62$\times10^{-18}$& 0.80  \\
\ion{O}{iv} &  8$\alpha$ & 1.7363 & 1.32$\times10^{-17}$  &  5.97$\times10^{-18}$& 0.74  \\
\hline
\ion{O}{iii}&  5$\alpha$ & 0.8262 & 6.03$\times10^{-17}$  &  4.57$\times10^{-18}$& 0.25  \\
\ion{O}{iii}&  6$\alpha$ & 1.3727 & 2.65$\times10^{-17}$  &  8.95$\times10^{-18}$& 0.25  \\
\ion{O}{iii}&  7$\alpha$ & 2.1149 & 1.28$\times10^{-17}$  &  6.78$\times10^{-18}$& 0.76  \\
\hline
\multicolumn{2}{l|}{[\ion{O}{iii}]}
                         & 0.5007 & 9.32$\times10^{-14}$  &  6.30$\times10^{-16}$&   \\
\multicolumn{2}{l|}{[\ion{O}{iii}]$^{**}$}
                         & 0.5007 & 1.11$\times10^{-12}$  &  7.52$\times10^{-15}$&   \\
\hline
\multicolumn{6}{l}
    {{\large\strut}$^*$ Including 11$\beta$ line that is overlapping with 9$\alpha$}\\
\multicolumn{6}{l}
    {{\large\strut}$^{**}$ Including the pre-shock photoionized region contribution}\\
\end{tabular}
\end{center}
\emph{Note}.
         Assumed cloud area is $1\times10^{33}$~cm (size about 0\farcs6).
	 Reddening was applied
         using~\protect\citet{RedLaw77} reddening curve
         and $E(B-V)=1.5$~\protect\citep{CasAFesen96}.
         The brightest knots are a factor of 3--10 more intense 
         in the [\ion{O}{iii}] 5007~\AA\ line than the values
         given in the table.
         {The $f_{\rm s}$ values were determined from simulated
         spectra having 200~km/s Doppler line width at the half maximum
         level (FWHM), that is equal to the optical line widths
         observed in the Cas~A knots \protect\citep{Bergh71}; 
         see also Section~\ref{SecLineStr}.}
\end{table}

\subsection{Cold photoionized region}

The ions in the photoionized region before the shock front are rapidly
recombining because of their low temperature. 
This results in rather bright recombination lines,
with fluxes proportional to the distance $d$ between
the ionization front and the shock wave.

The plain-parallel SD95 model does not determine this distance.
Assumption of small optical depth of material between the shock wave
and the ionization front results in restriction of $d<10^{17}$~cm
for relevant cloud models for preshock ion number density of
100~cm$^{-3}$ (illustrated on the Figure 12 of SD95).
Below we take conservatively $d=10^{16}$~cm, keeping in mind that the
line fluxes scale linearly with it.

The cloud model also do not determine exact temperature in the
photoionized region. We take $T_{\rm e}=1\times10^3$~K as a reference
value, but \citet{Dopita84} mention that it may be as low as 100~K.
If the real temperature in the region is less than our assumed value,
recombination line emissivities are respectively higher 
(see Section~\ref{SecTdiag} below for quantitative dependences).

\begin{figure}
\begin{center}
\centerline{
    \rotatebox{270}{
       \includegraphics[height=0.95\linewidth]{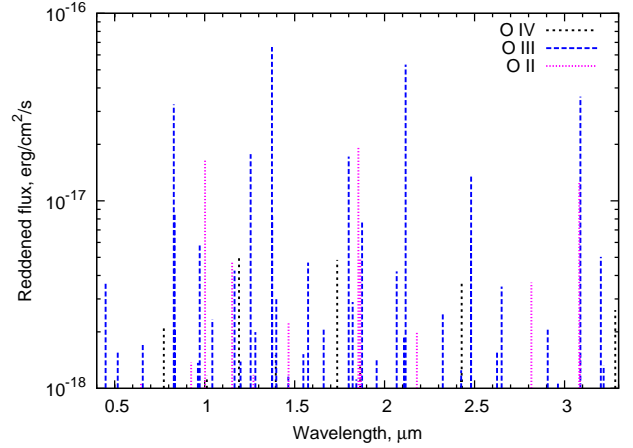}
                   }
           }
\caption{Recombination line $l$-summed reddened fluxes expected from
         the photoionized region before the shock front
         computed following the~SD95 ionization front model.
         Details are as in note to Table~\protect\ref{TabFluxSW}.
         Photoionized region thickness is $10^{16}$~cm.
         }
   \label{FigPI}
  \end{center}
\end{figure}

Predicted oxygen line fluxes for the SD95 model OP200
with shock speed of 200~km/s and 
total preshock ion number density 100~cm$^{-3}$
are shown on Figure~\ref{FigPI} and given in Table~\ref{TabFluxPI}.
It is seen that even with our quite conservative assumptions
resulting line fluxes are still somewhat stronger than
from the shock front.
Indeed, the predicted line brightest components are on the level
of almost {half a percent} of the [\ion{O}{iii}] 5007~\AA\ line. 

\begin{table}
\caption{Brightest recombination line $l$-summed 
         emitted and reddened fluxes $I$ and $F$
         from the photoionized region before the shock front.
         computed following the SD95 model.
         Details are as in note to Table~\protect\ref{TabFluxSW}.
         Photoionized region thickness is $10^{16}$~cm.
         The last column contains the flux part in the strongest
         unresolved line component $f_{\rm s}$ for FWHM of 200~km/s
         {(see Note to Table~\protect\ref{TabFluxSW})}.
         The wavelength $\lambda$ corresponds to the strongest line component.
\label{TabFluxPI}
}
\begin{center}
\begin{tabular}{lr|llll}
\hline\hline
Ion & Line & $\lambda$, $\mu$m &  $I$, erg/cm$^2$/s & $F$, erg/cm$^2$/s&  $f_{\rm s}$ \\
\hline
\ion{O}{iii}&  5$\alpha$  & 0.8262  &  4.13$\times10^{-16}$ & 3.26$\times10^{-17}$& 0.26 \\
\ion{O}{iii}&  6$\beta$   & 0.8327  &  1.05$\times10^{-16}$ & 8.48$\times10^{-18}$& 0.31 \\
\ion{O}{iii}&  7$\beta$   & 1.2552  &  6.09$\times10^{-17}$ & 1.77$\times10^{-17}$& 0.73 \\
\ion{O}{iii}&  6$\alpha$  & 1.3727  &  1.95$\times10^{-16}$ & 6.61$\times10^{-17}$& 0.27 \\
\ion{O}{iii}&  8$\beta$   & 1.7998  &  3.69$\times10^{-17}$ & 1.72$\times10^{-17}$& 0.86 \\
\ion{O}{iii}&  7$\alpha$  & 2.1149  &  1.00$\times10^{-16}$ & 5.32$\times10^{-17}$& 0.82 \\
\ion{O}{iii}&  9$\beta$   & 2.4804  &  2.32$\times10^{-17}$ & 1.36$\times10^{-17}$& 1.32$^*$ \\
\ion{O}{iii}&  8$\alpha$  & 3.0871  &  5.50$\times10^{-17}$ & 3.59$\times10^{-17}$& 0.91 \\
\hline
\ion{O}{ii} &  5$\alpha$  & 1.8645  &  4.03$\times10^{-17}$ & 1.94$\times10^{-17}$& 0.24  \\
\ion{O}{ii} &  6$\alpha$  & 3.0922  &  1.90$\times10^{-17}$ & 1.24$\times10^{-17}$& 0.59  \\
\hline
\multicolumn{6}{l}{{\large\strut}$^*$ Including 10$\gamma$ line that is overlapping with 9$\beta$}
\end{tabular}
\end{center}
\end{table}

{If they would be just a factor of several stronger,}
the lines would have been detected
in spectroscopic observations of the fast-moving knots
(see discussion on existing observation limits in
Section~\ref{SecLimits} below).

Our estimate of the observational limit 
on the \ion{O}{iii} 5$\alpha$ recombination line flux at 0.83~$\mu$m
based on the data by~\citet{CasAFesen96}
of $0.005\times F$(5007~\AA)
may be transformed into a joint constraint on the
thickness of the photoionized region $d$,
its temperature $T_{\rm e}$ and total ion density $n_{\rm t}$:
$$
 \left(\frac{n_{\rm t}}{100\,{\rm cm^{-3}}}\right)^2\,
 \left(\frac{10^3\,{\rm K}}{T_{\rm e}}\right)  \,
 \left(\frac{d}{10^{16}\,{\rm cm}}\right)
    < 7,
$$
assuming $E(B-V)=1.5$ and
\element[3+]{O} ionic abundance after the ionization front of 0.6.

On this example it is easy to see that detection of the lines
or even tight upper limits will allow to perform detailed
observational tests of the SD95 and other FMK models.

\subsection{Separating pre- and post-shock spectral lines}

Observationally, the lines formed in the regions before and after
the shock wave will be distinguishable because of two effects.
Firstly,
the lines arising in the photoionized region yet dynamically undisturbed
by the shock wave should be intrinsically very narrow,
but the lines arising in the post-shock region are
expected to be Doppler-broadened by turbulence.

Secondly, the spectral lines arising in the post-shock region
should be shifted with respect to the pre-shock lines.
The relative Doppler shift between these two line families appears 
due to motion of the post-shock gas
relative to the initial knot speed induced by the shock wave.
In the strong-shock limit this velocity difference 
equals $3/4\, \upsilon_{\rm shock}$ for specific heat ratio $\gamma=5/3$ 
\citep{ZeldRaiz66}, i.e. up to 150~km/s in the SD95 model,
but the relative Doppler shift will also depend on the angle between
the shock front and the direction towards observer.

Both collisionally-excited and recombination lines should display
these features, but for the former group such differences should be
easier to detect because of higher line intensities.

The observations with higher spectral resolution 
($\lambda/\Delta\lambda\la20\,000$) could also result
in detection of several narrow components forming a line.
They could arise from the individual vortice
or velocity component emission, if there is only one or a few
of them dominating the plasma motions after the shock,
like observed in laser experiments simulating cloud-shock interaction 
in supernova remnants \citep{CloudCrushExp03}.
The resulting spectral line profile might be quite complicated 
(e.g., \citet{KKHP-SK}), 
but such observations will provide data on 
the post-shock dynamics of the fast-moving knot
that is presently poorly constrained from observations.

\subsection{Existing observational limits}
\label{SecLimits}

As it is seen from Figures~\ref{FigSpeSD95} and~\ref{FigPI},
the brightest recombination lines are expected to lie in the wavelength range
between 0.7 and 3~$\mu$m. Several detailed spectroscopic investigations
have already been performed in this range and could have found these
lines, provided that they had low enough detection limits.
Below we discuss such existing optical and near-infrared limits
on the line fluxes and compare them with the SD95 model predictions.

Optical spectra of fast-moving knots with wide spectral coverage
were obtained, e.g., by \citet{CheKirAbund} and \citet{CasAFesen96}. 
According to the authors, the detection limits
around 0.75--0.85~$\mu$m are about 300 times less than the [\ion{O}{iii}]
5007~\AA\ line flux for the brightest observed features.

Our results give flux ratios of the reddened 5007~\AA\ line
to the brightest components of the \ion{O}{vi} 8$\alpha$,
\ion{O}{v} 7$\alpha$ and \ion{O}{iii} 5$\alpha$ lines
of about 300--500 
(see Figures~\ref{FigSpeSD95} and~\ref{FigPI} 
 and Tables~\ref{TabFluxSW} and \ref{TabFluxPI}).
These values show that the model is on the boundary of consistency
with the observational results.
They also imply that the oxygen recombination line detection
should be possible, provided that 
the physics in the post-shock and pre-shock regions corresponds
to the one described by the SD95 model.

Recently, near-infrared (0.95--2.5~$\mu$m) 
spectra of Cas~A fast-moving knots, obtained
at the 2.4~m Hiltner telescope, were published by \citet{CasANIR01}.
It is more difficult to compare their detection limits with our
predictions, as there is only one oxygen \ion{O}{i} line blend
present around 1.129~$\mu$m.
It arises in transitions between excited states of neutral oxygen
and is also blended with [\ion{S}{i}] line. 
If we assume that the overlapping [\ion{S}{i}] line emission
is negligible, our estimate of the observational detection limit
corresponds to about 1/100 of the reddened optical [\ion{O}{iii}]
line flux.
This value is a factor of several higher than the predicted fluxes
of the \ion{O}{vi} 9$\alpha$, \ion{O}{v} 8$\alpha$ 
and \ion{O}{iii} 6$\alpha$ lines, also showing feasibility of
the line detection.

\section{Individual line substructure}
\label{SecLineStr}

\begin{figure*}
\begin{center}
\centerline{
    \rotatebox{270}{
        \includegraphics[height=0.47\linewidth]{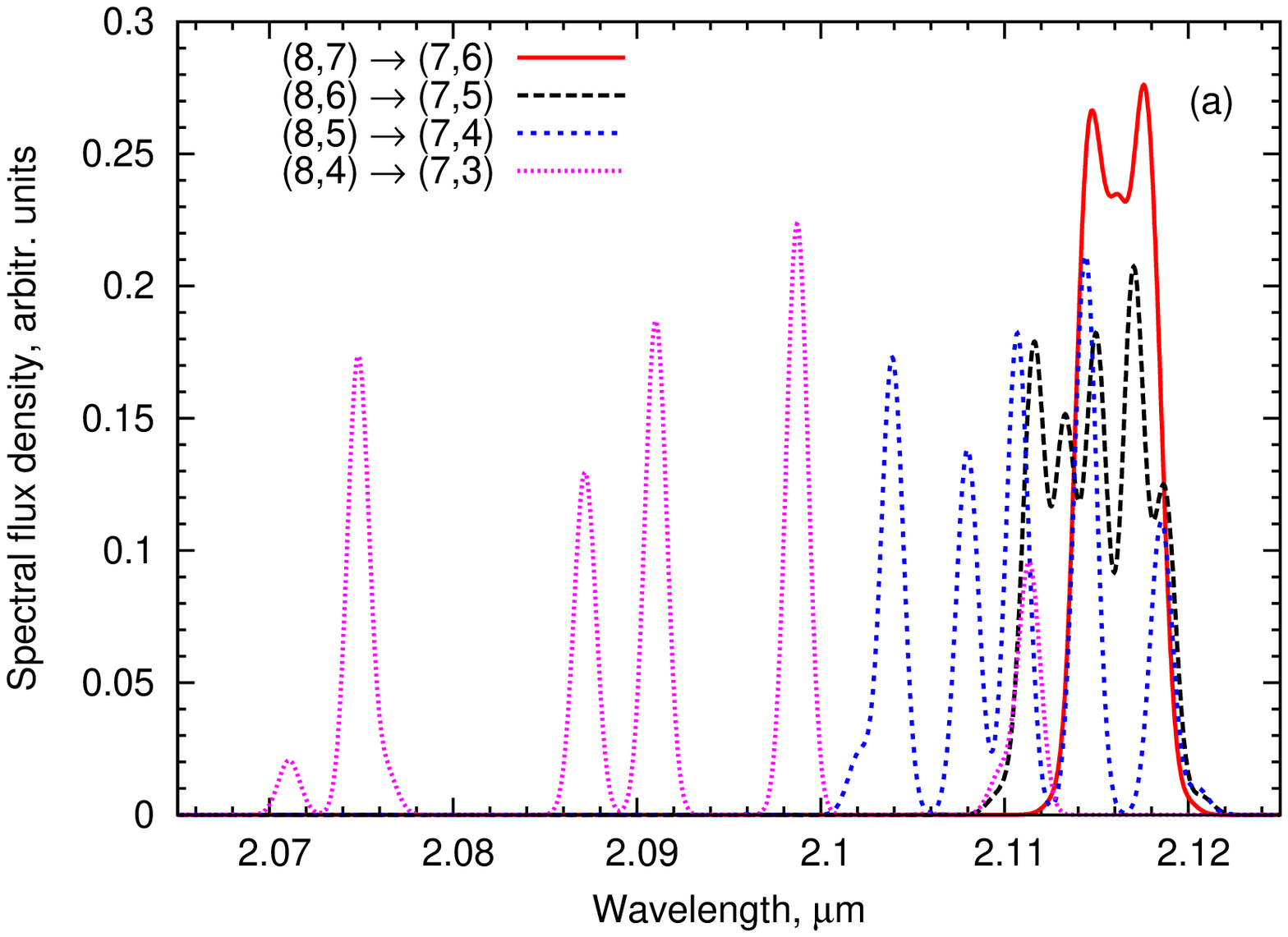}
                   }
    \rotatebox{270}{
        \includegraphics[height=0.47\linewidth]{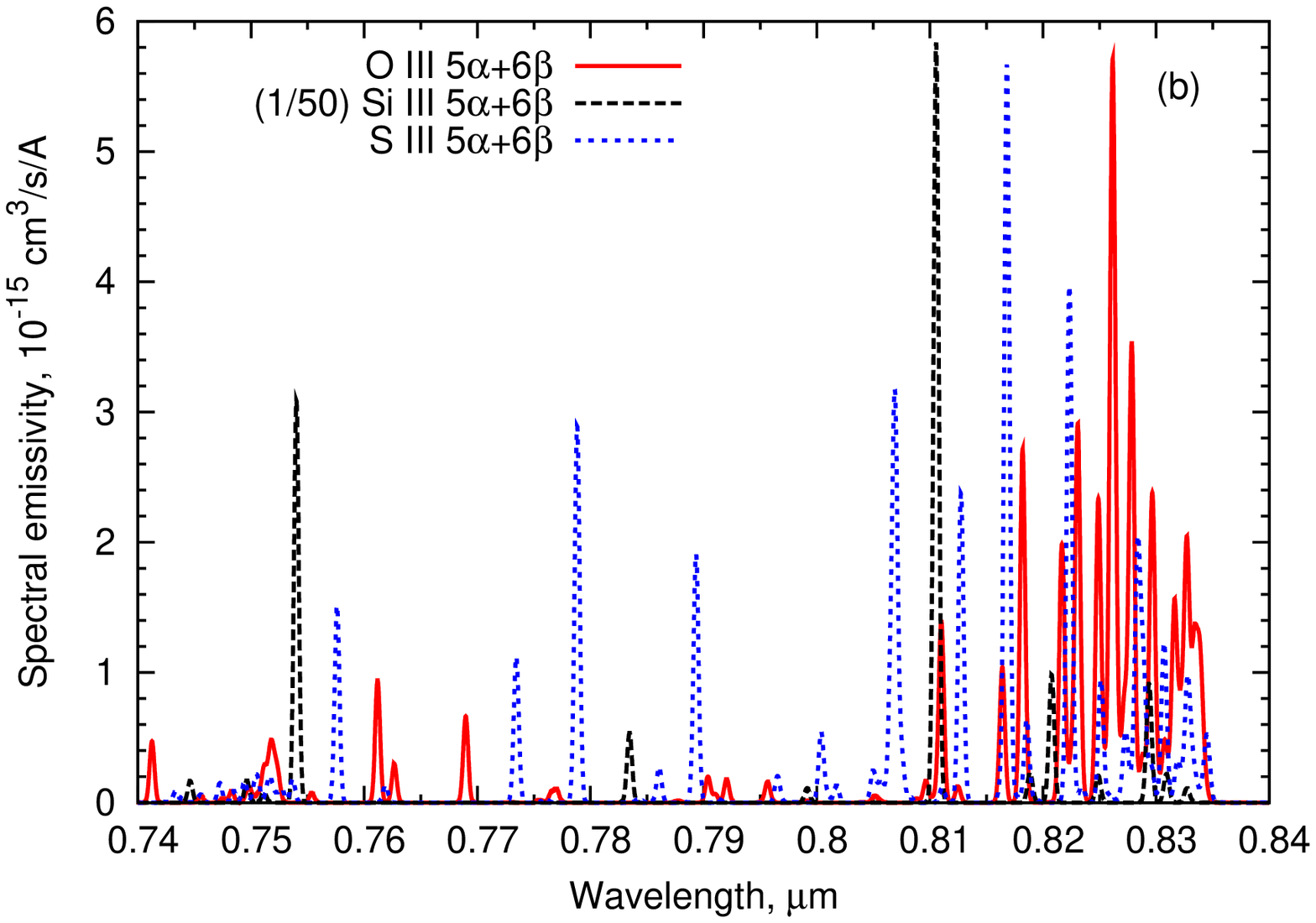}
                   }
           }
\centerline{
    \rotatebox{270}{
        \includegraphics[height=0.47\linewidth]{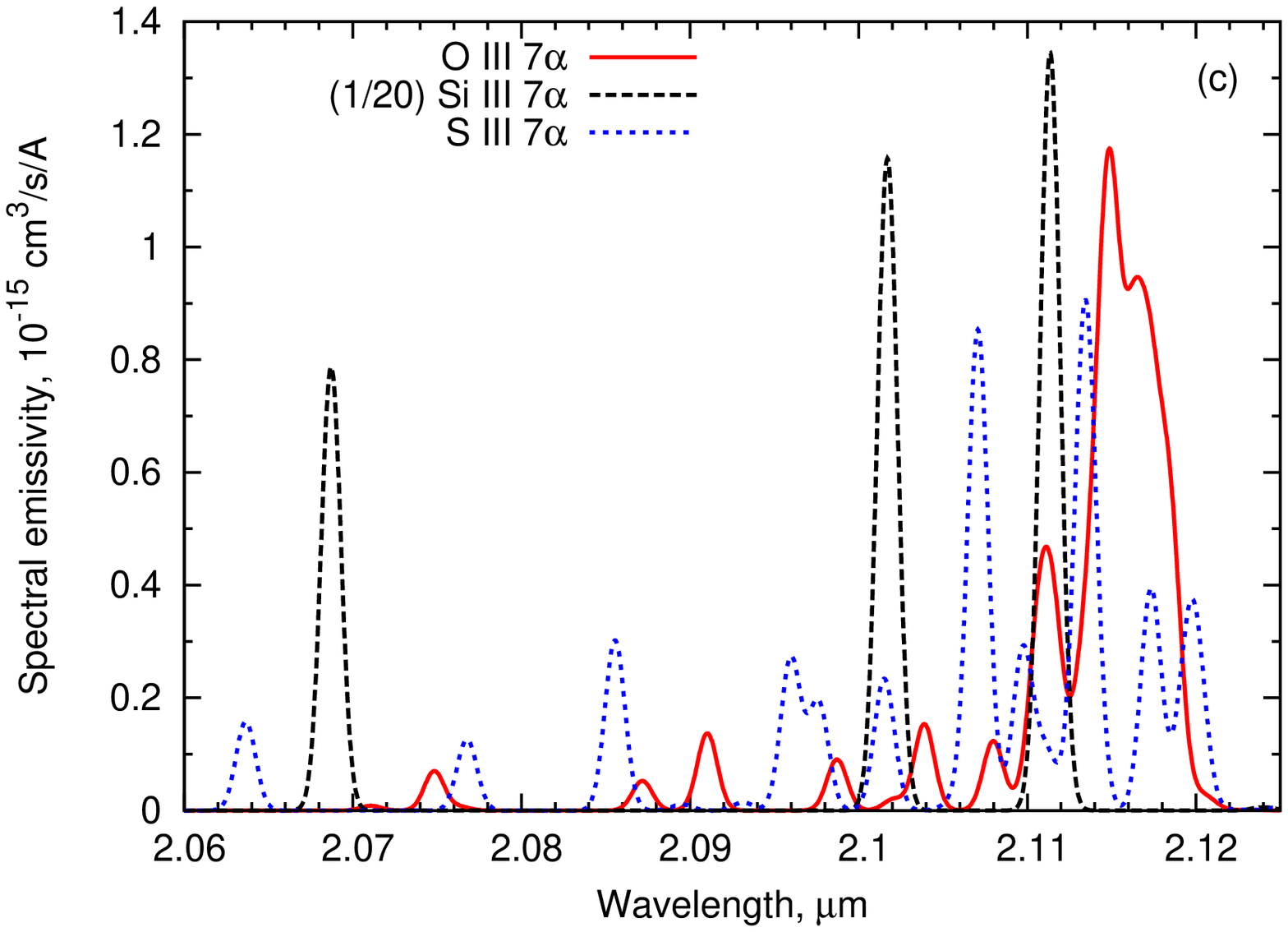}
                   }
    \rotatebox{270}{
        \includegraphics[height=0.47\linewidth]{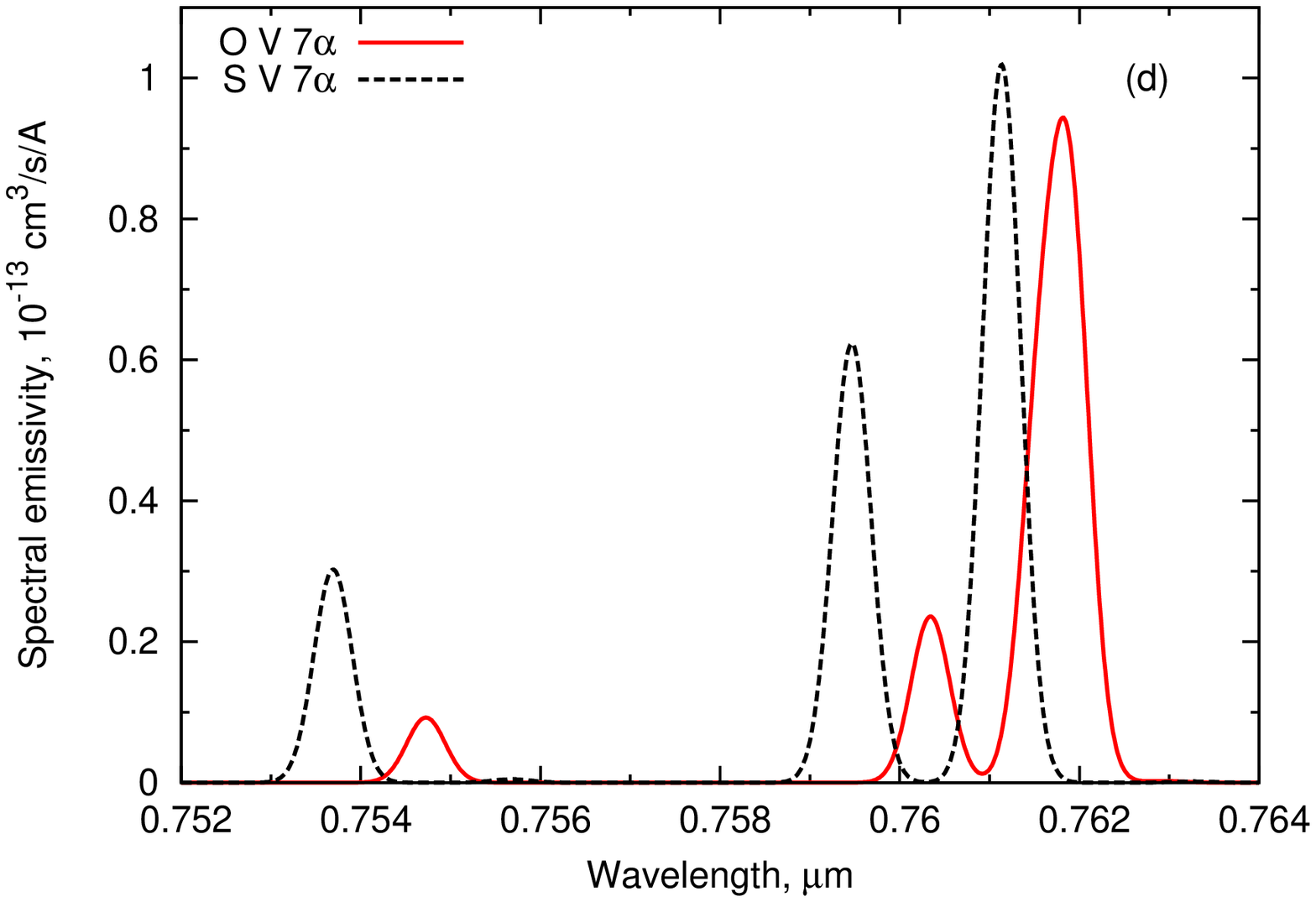}
                   }
           }
\caption{Fine structure of recombination lines. Line Doppler width 
         at half maximum of 200~km/s is assumed.
         (a): \ion{O}{iii} $7\alpha$ line $(8,l)\to(7,l-1)$ component structure
            for $l\ge 4$.
            Sum of all components is set to be equal for each $l$.
            Emissivities of the $(8,l)\to(7,l+1)$ components are much weaker and
            structure of resulting lines is not shown.
         (b),(c): Spectral emissivity d$\varepsilon/$d$\lambda$, 
            cm$^3$/s/\AA\ of the simulated 
            \ion{O}{iii}, \ion{Si}{iii} and \ion{S}{iii} 
            $5\alpha+6\beta$ and $7\alpha$ line structure
            for low-density case and temperature $T_{\rm e}=2\times10^4$~K.
            Note that \ion{Si}{iii} lines have
            been scaled down to fit in the same scale.
         (d): Spectral emissivity for the simulated
            \ion{O}{v} and \ion{S}{v} $7\alpha$ line structure
            in the same conditions. Absence of $K$-splitting
            results in much simpler profiles.
         }
   \label{FigLineFS}
  \end{center}
\end{figure*}

In the non-relativistic hydrogenic approximation, level energies
depend only on 
{the ion charge and the principal quantum number $n$ of the level.
Though, this approximation is not fully applicable for the description
of the spectral lines corresponding to transitions between
the levels with $n\approx10$, as other effects start
playing a role.} 
They arise due to energy level shifts, described in 
Appendix~\ref{SecQuaDef},
resulting in the ``fine'' structure appearing in the spectral lines.
{As a result, these effects help}
to distinguish lines of different elements having equal
ionization stages {and initial and final $n$'s}.

We calculated the line $l$-- and $K$--substructure
following the method outlined in  Appendix~\ref{SecTLineStr}
($K$ is the quantum number used to characterize
additional interaction of the highly-excited electron orbital momentum
with the total angular momentum of other electrons,
present in \ion{O}{ii}, \ion{O}{iii}, \ion{S}{ii} and \ion{S}{iii} lines).
The results are shown on Figure~\ref{FigLineFS} for several most
interesting cases, assuming recombination lines having widths around 200~km/s 
(full width at half maximum level, FWHM), 
of the same order as the observed widths of forbidden
optical lines~\citep{Bergh71}.

It is easily seen that high-$l$ levels
have both smaller quantum defects and smaller $K$-splitting.
Complicated line structure in case of presence of the $K$-splitting
results in weaker individual line fluxes making them more difficult
to detect.

One should also keep in mind that spectral lines arising in transitions
between lower-$n$ levels have higher $K$- and $l$-splittings.
Therefore the line components have larger separations
and the amount of individual components increase even more
with each of them being respectivly less intense.
As a bottomline, the near-infrared lines are more promising
for the first detection than the optical ones, as they have less
substructure (compare panels (b) and (c) on Figure~\ref{FigLineFS}).

The substructure of the 6$\alpha$ recombination lines
of \ion{Mg}{i}, \ion{Al}{i} and \ion{Si}{i}
have been observed in solar spectra near
12~$\mu$m and explained using similar, although somewhat
more elaborate, theoretical description
\citep{Sun12mkm,ChangSun12mkm,Chang84}.
{We have used these observations as the test cases for our line
substructure computation code.}

In Tables~5--15, available as an electronic supplement at the CDS,
we give \ion{O}{ii} -- \ion{O}{vi}, 
\ion{S}{ii} -- \ion{S}{v}, \ion{Si}{ii} and \ion{Si}{iii}
recombination line component vacuum wavelengths, emissivities
and relative intensities in the $(nlK)\to(n'l'K')$ resolution 
for several temperatures 
(lg$(T_{\rm e}, {\rm K}) =$3.0, 3.5, 4.0, 4.5 and 5.0)
in the low-density limit.
First seven columns characterize the quantum numbers of the transition
(total recombining ion electronic angular momentum $J_{\rm c}$
 and highly-excited electron quantum numbers 
 before and after the transition $n$, $l$, $K$, $n'$, $l'$, $K'$),
 Column 8 gives the line component wavelength in microns,
 next columns in pairs state line component emissivity in cm$^3$/s
 and intensity ratio of this component with respect to the 
 $K$- and $l$-summed emissivity 
 $\varepsilon(J_{\rm c},nlK,n'l'K')/\varepsilon(n,n')$.
Only the lines most likely to be detected are given in these tables,
selected by the following parameters -- only $\alpha$, $\beta$ and $\gamma$
lines having wavelengths between 0.3 and 5.0~$\mu$m.

Tables 16--23, also available at the CDS, contain similar
information on predicted oxygen line component fluxes 
in the SD95 200~km/s shock model.
The fluxes from the cooling and photoionized regions are given
separately in Tables~16--20 (for \ion{O}{ii} -- \ion{O}{vi})
and 21--23 (for \ion{O}{ii} -- \ion{O}{iv}), respectively.
First eight columns again contain the quantum numbers characterizing 
the transition and the line wavelength, Column~9 lists line
component flux in erg/cm$^2$/s, Column~10 contains intensity ratio
of this component with respect to the $K$- and $l$-summed emissivity.

As discussed in Appendix~\ref{SecQuaDef},
the input atomic data are precise to about 10\%
and we cannot expect better precision of the 
resulting line wavelength differences from the hydrogenic values.

\begin{figure}
\begin{center}
\centerline{
    \rotatebox{270}{
       \includegraphics[height=0.95\linewidth]{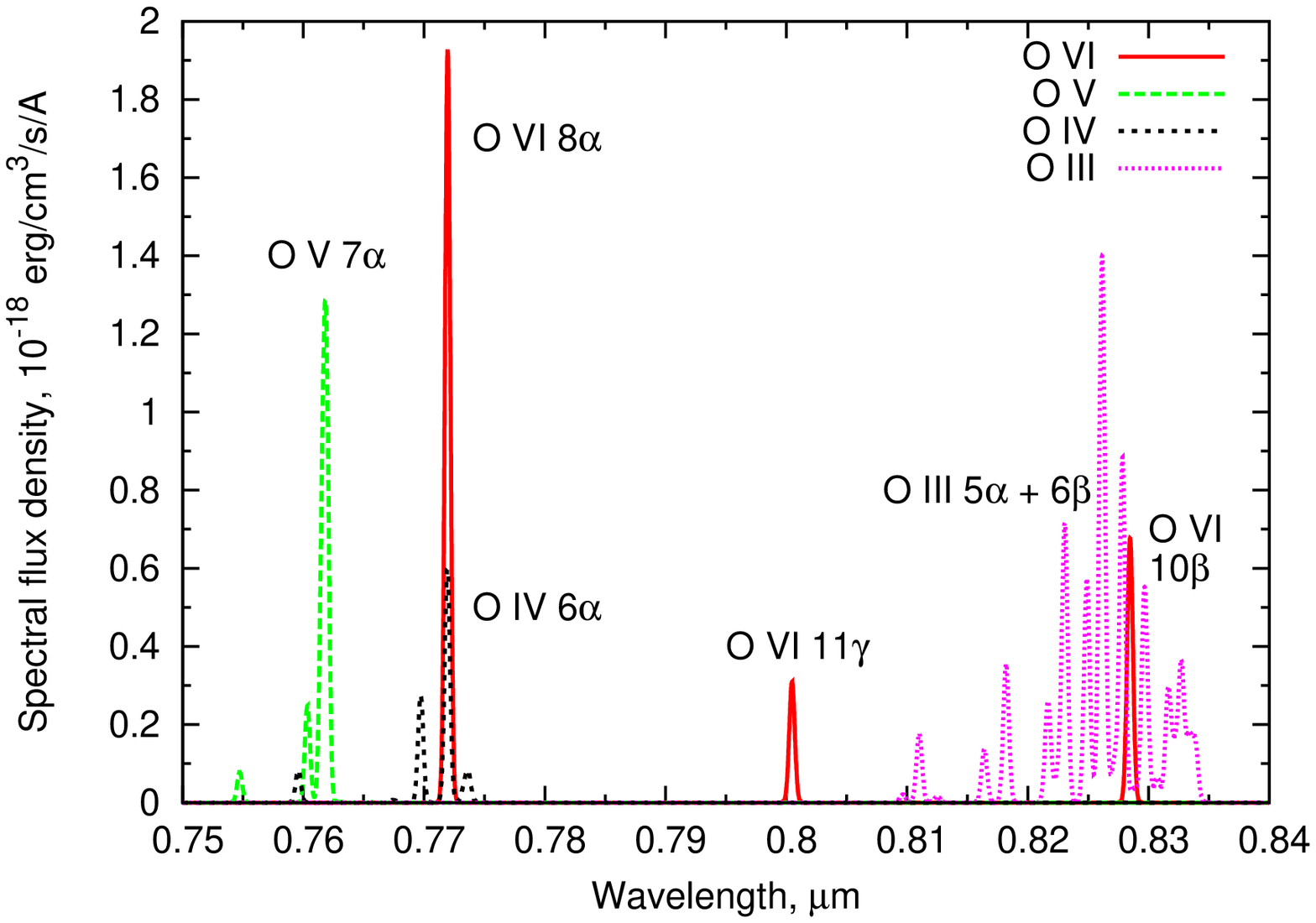}
                   }
           }
\centerline{
    \rotatebox{270}{
       \includegraphics[height=0.95\linewidth]{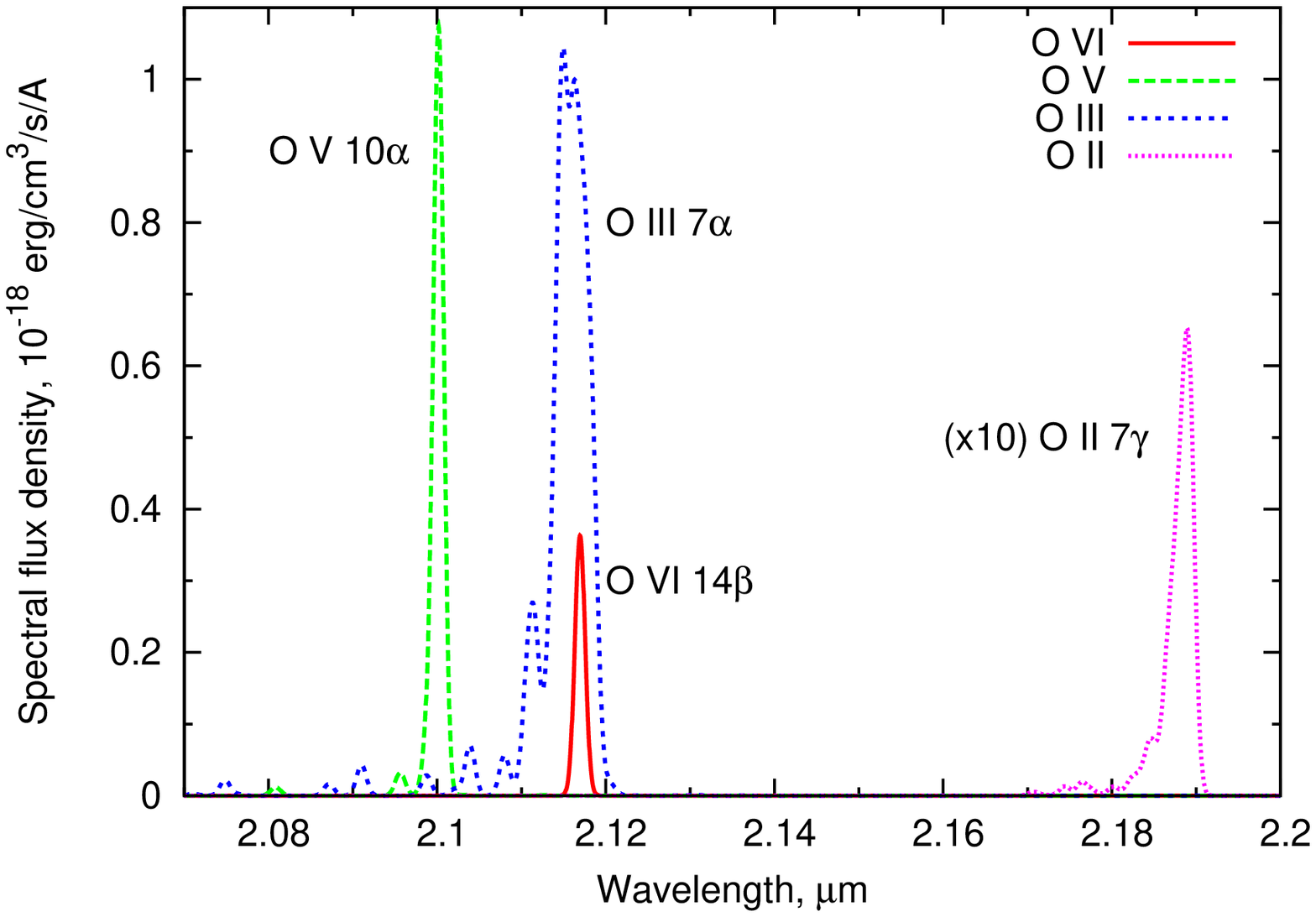}
                   }
           }
\caption{Model spectra near 0.8 and 2.1~$\mu$m 
         containing recombination lines of \ion{O}{ii}--\ion{O}{vi} 
         computed based on the SD95 200~km/s shock model.
         Contributions of cooling and photoionized regions are summed up.
         It is seen that at longer wavelengths the substructure
         is less influencing the line profiles.
         Note that \ion{O}{ii} $7\gamma$ line intensity is multiplied by 10
         for illustration purposes. Line FWHM of 200~km/s is assumed.
         }
   \label{FigSD95tspe}
  \end{center}
\end{figure}

Sample region from the total resulting model spectrum is shown
on Figure~\ref{FigSD95tspe}. It shows variety of the spectral shapes
as well as illustrates the diminishing of the peak intensity due
to $K$-splitting for \ion{O}{iii} lines.

\bigskip 

The line substructure change with density and temperature because
of changes in the relative populations of the $l$-states.
At low temperatures the main population mechanism is the
radiative recombination, which is populating high-$l$ levels
relatively efficient.
When the electron density increases, $l$-redistribution modifies
high-$l$ populations for $n$ greater than about 20, but for $n<15$
the changes are generally smaller.

At higher temperatures, DR populates relatively lower $l$ states
having higher quantum defects and low probabilities of $\alpha$-line
emission.
Therefore a recombination line is split into many components and
its emissivity is relatively low.
In this case, $l$-redistribution mostly shift the recombined electrons
to higher-$l$ states, simplifying the line profile
and significantly increasing its emissivity.

\begin{figure}
\begin{center}
\centerline{
    \rotatebox{270}{
       \includegraphics[height=0.95\linewidth]{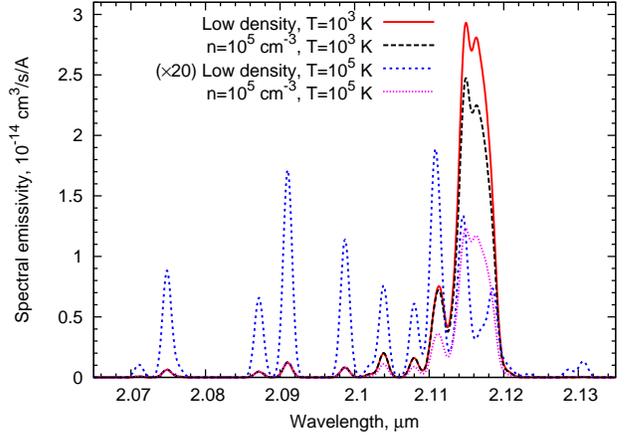}
                   }
           }
\caption{Variation of the fine structure of the 
         \ion{O}{iii} $7\alpha$ line at 2.1~$\mu$m
         with temperature and density. 
         Line FWHM of 200~km/s is assumed.
         Note that the low-density curve for $T_{\rm e}=10^5$~K
         is multiplied by 20 to be visible on this scale.
         }
   \label{FigOIIInet}
  \end{center}
\end{figure}

On Figure~\ref{FigOIIInet} we show the change of the \ion{O}{iii}
$7\alpha$ line fine structure with temperature and density,
illustrating the described effects.

Another effect arising at low temperatures and electron densities
 -- lower populations of excited $J_{\rm c}$ core states.
For \ion{O}{ii}, \ion{O}{iii}, \ion{S}{ii} and \ion{S}{iii} lines 
this results in damping of the
$K$-splitting, as it arises only from recombination of excited ions.
All the flux from $(nlK)\to(n'l'K')$ components in this case
will be redistributed into the central $(nl)\to(n'l')$ components
and only the $l$-splitting will remain.

\section{Plasma diagnostics using recombination lines}
\label{SecModResults}

\subsection{Temperature diagnostics}
\label{SecTdiag}

\begin{figure}
\begin{center}
\centerline{
    \rotatebox{270}{
        \includegraphics[height=0.95\linewidth]{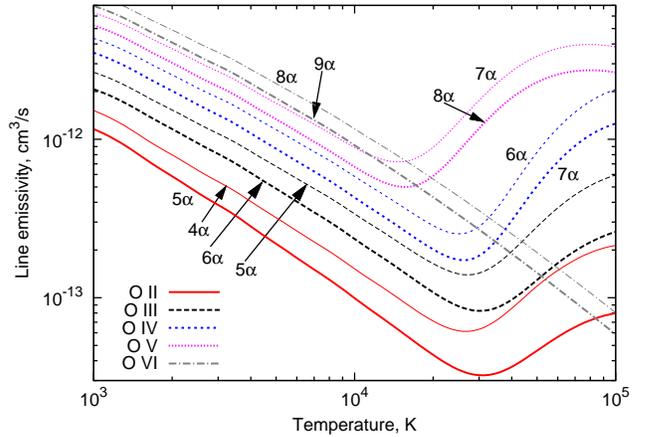}
                   }
           }
\caption{Low-density $l$-summed
         emissivities of oxygen ion optical and near-infrared 
         recombination lines. Different colors represent different ions,
         plotted line thickness increase with the spectral line wavelength.
         }
   \label{FigOepsT}
  \end{center}
\end{figure}

Line emissivity dependence on temperature is shown for several
bright optical and near-infrared oxygen recombination lines
in the low-density limit on Figure~\ref{FigOepsT}.
Corresponding line wavelengths are given in Table~\ref{TabRLlam}
above.
This figure allows to compute the recombination line fluxes
for other models of the multi-temperature plasma with emission
measure distribution different from the discussed~SD95 model.

The curves on Figure~\ref{FigOepsT} has two distinct regions.
At low temperatures the emissivity is determined by the radiative
recombination and changes smoothly from ion to ion.
At temperatures higher than several tens of thousands Kelvin
the dielectronic recombination starts to dominate the recombination rates
({except for the \ion{O}{vi} lines due to absence at low temperatures
of dielectronic recombination channels in He-like ion \element[6+]{O}})
and line emissivities of different ions
start being determined by the lowest excited states of the recombining ion,
as described in Appendix~\ref{SecDR}.
The temperature of the strong emissivity rise is proportional
to the typical energy of such excited states. 
Amplitude of the rise is proportional
to the lowest excited state decay rate.

For the discussed~SD95 200~km/s shock model the high-temperature
region where the DR dominates is important only for \ion{O}{v} lines,
but it is easy to imagine other ionic abundance distributions where
it will affect also ions in lower ionization stages.

\begin{figure}
\begin{center}
\centerline{
    \rotatebox{270}{
        \includegraphics[height=0.95\linewidth]{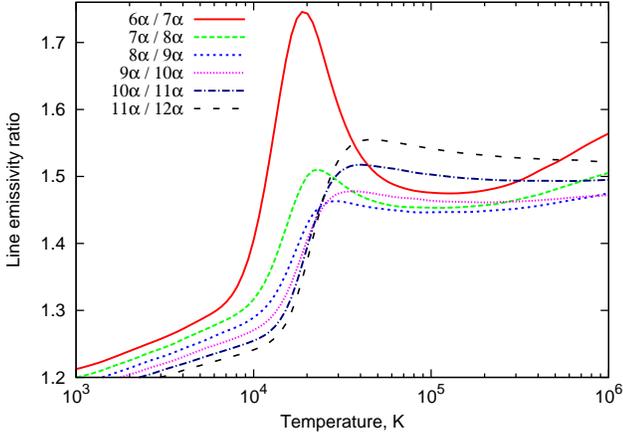}
                   }
           }
\caption{Low-density $l$-summed \ion{O}{v} recombination line 
         emissivity ratios as functions of electron temperature.
         }
   \label{FigOratiosT}
  \end{center}
\end{figure}

Dependence of individual \ion{O}{v} recombination $\alpha$-line ratios
on temperature is shown on Figure~\ref{FigOratiosT}.
Again, the two distinct regions are seen at low and high temperatures
corresponding to RR- and DR-dominated recombination.

It is easy to notice that in the intermediate temperatures,
the lines corresponding to transitions at lowest $n$'s become
relatively brighter.
Reason for such dependence, especially clearly visible on the
$6\alpha/7\alpha$ line ratio, is the following. As follows from
Appendix~\ref{SecDR}, at low temperatures the dielectronic recombination rate
is damped by the factor $\exp\left(-{\mathcal E}/{k T_{\rm e}}\right)$.
For the low $n$-levels the doubly excited state energy $\mathcal E$ may
be significantly lower than the core excitation energy and at low
temperatures this makes a difference and the dielectronic recombination
populates mostly low-$n$ levels,
increasing emissivities of low-$n$ recombination lines.

The described effect is most pronounced for recombination lines
of \ion{O}{v}, \ion{Si}{iii} and \ion{S}{v}, although in all cases the most
temperature-sensitive line -- 
the $\alpha$-line formed by the transition from the lowest
level, onto which DR is possible --
is situated around 2000--3000~\AA.
Though, as can be seen from Figure~\ref{FigOratiosT}, also the next line
situated for these ions around 4500--5000~\AA, can also be a useful
diagnostical tool, being relatively much brighter in a narrow
temperature interval.

\subsection{Recombination lines as density diagnostics}
\label{SecRLdens}

Recombination line flux ratios allow in principle
to determine also density of the emitting region, as their dependences
on density and temperature differently affect the relative line fluxes.

\begin{figure}
\begin{center}
\centerline{
    \rotatebox{270}{
       \includegraphics[height=0.95\linewidth]{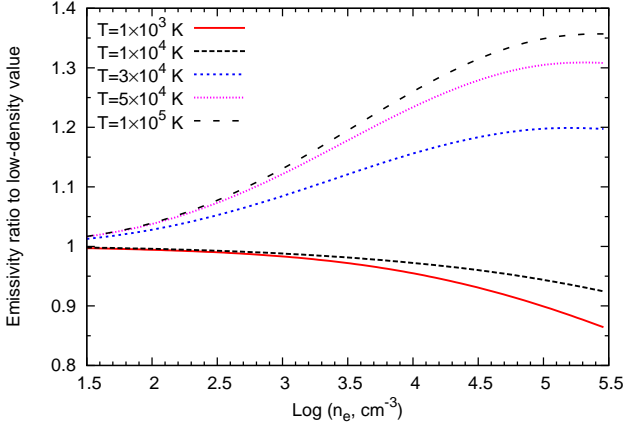}
                   }
           }
\caption{Dependence of \ion{O}{v} $7\alpha$ line emissivity on electron
         density for different electron temperatures $T_{\rm e}$.
         Values are normalized to the low-density emissivity.
         }
   \label{FigEmneT}
  \end{center}
\end{figure}

As an example, on Figure~\ref{FigEmneT} we plot the \ion{O}{v}
$7\alpha$ line emissivity as function of the electron density for
different temperatures between $10^3$ and $10^5$~K.
For better representation, emissivity ratios to their low-density
values are shown.
Clearly, the density dependences are weak and can probably
be neglected in the first stage of the qualitative analysis,
especially at low temperatures.

Two distinct regions are seen once again. 
At low temperatures where only radiative recombination
determines the level populations, emissivities decrease
with density, but at higher temperatures they increase.
The difference is explained by different initial populations
of the levels.

Dielectronic recombination predominantly populates states
with $l\la10$ (see, e.g., Appendix~\ref{SecDR}).
Collisions in this case mostly transfer recombined electrons
to higher $l$'s,
increasing the probability of $\Delta n=1$ transitions and,
therefore, $\alpha$-line emissivities.

Radiative recombination, especially at low temperatures, populates
high-$l$ states much more efficiently. In this case, populations 
of low-$l$ levels relatively increase as a result of
the $l$-changing collisions, and recombination line emissivities
somewhat decrease.

\begin{figure}
\begin{center}
\centerline{
    \rotatebox{270}{
       \includegraphics[height=0.95\linewidth]{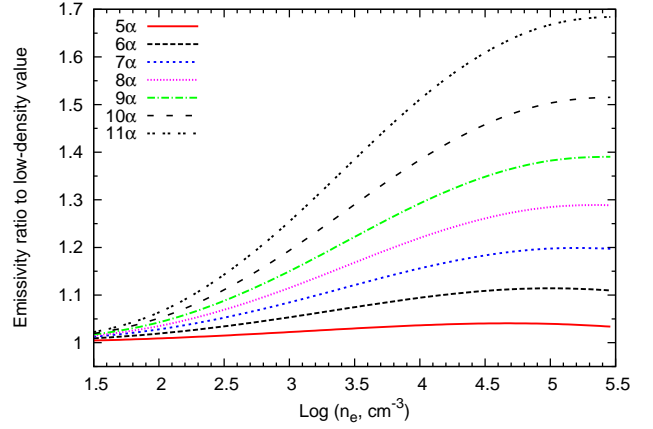}
                   }
           }
\caption{Increase of \ion{O}{v} $\alpha$-line emissivities with electron
         density at temperature $T_{\rm e}=3\times10^4$~K.
         }
   \label{FigEmnen}
  \end{center}
\end{figure}

On Figure~\ref{FigEmnen} we show the \ion{O}{v} recombination line
emissivities relative to their low-density values as functions
of electron density at temperature 30\,000~K.
Lines arising in transitions between higher levels increase relatively
more, but in absolute terms the increase in emissivity is approximately
constant, determined by the $l$-redistribution on the high-$n$ levels.

\subsection{Recombination lines of other elements}

Our previous analysis mainly concerned the oxygen recombination
line emission, as the SD95 model
allows making quantitative predictions about their fluxes
and line ratios to the observed collisionally-excited lines.
For other elements, no model results for ionic abundance distribution
are available. 
Thus to predict their line fluxes it is necessary to solve a separate
problem of non-equilibrium plasma cooling and recombination
after the shock front for dense clouds of different compositions.
Such analysis is outside the scope of this paper.

Nevertheless, it seems valuable to provide data on the 
recombination line emissivities as functions of temperature
for the most typical elements. Then, from measured line ratios,
it will be possible to reconstruct characteristic
conditions in the emitting regions.
As two most typical examples for elements composing FMKs in Cas~A, 
we concentrate on the recombination lines of silicon and sulphur.
Approximate wavelengths of their recombination lines
are given in Table~\ref{TabRLlam} above.

\begin{figure}
\begin{center}
\centerline{
    \rotatebox{270}{
        \includegraphics[height=0.95\linewidth]{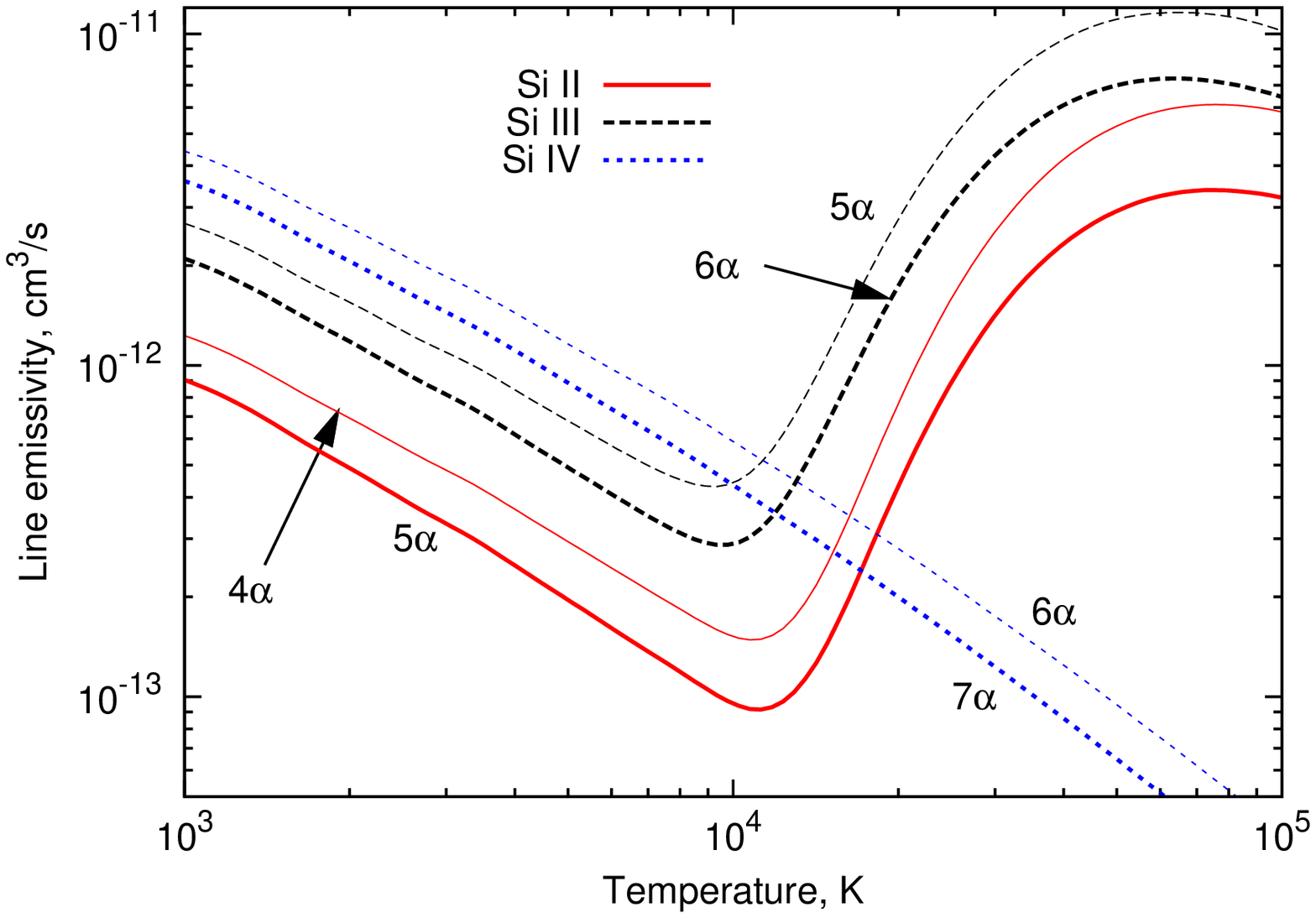}
                   }
           }
\centerline{
    \rotatebox{270}{
        \includegraphics[height=0.95\linewidth]{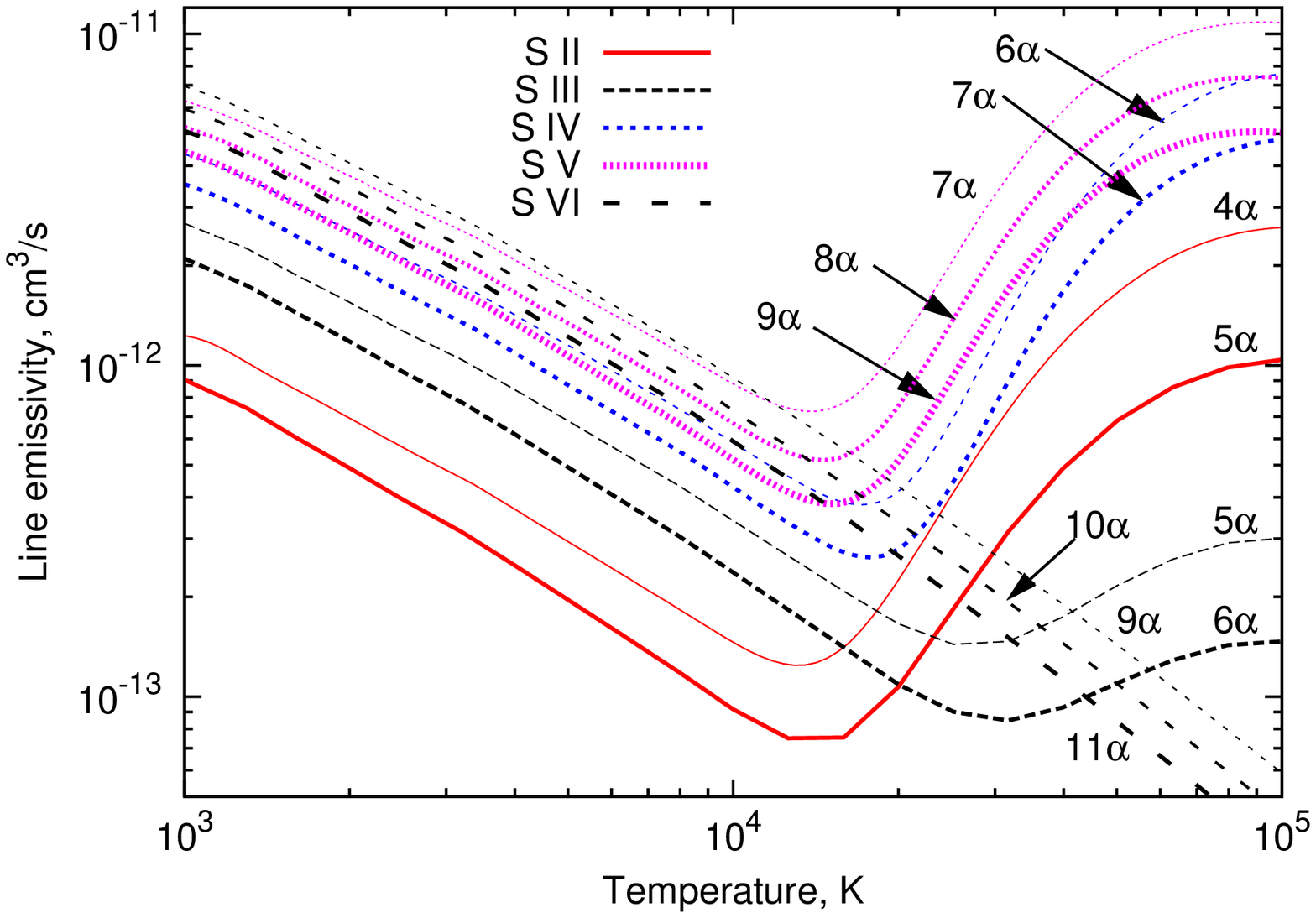}
                   }
           }
\caption{Low-density $l$-summed
         emissivities of several optical and near-infrared 
         silicon and sulphur recombination lines. 
         Different colors represent different ions,
         plotted line thickness increase with spectral line wavelength.
         }
   \label{FigSSieps}
  \end{center}
\end{figure}

Figure~\ref{FigSSieps} presents low-density emissivities of brightest
optical and near-infrared recombination lines of silicon and sulphur
ions, expected to exist in the fast-moving knots.
Note that \ion{Si}{iv} and \ion{S}{vi} lines are much weaker
than of the other ions at $T_{\rm e}>10^4$~K due to
weakness of dielectronic recombination channels at these temperatures.
Note also that the line emissivities in \ion{Si}{ii} and \ion{Si}{iii}
ions start to increase sharply already at about 12\,000~K.

Thus these ions are the most sensitive tracers of plasma
at temperatures between 15\,000 and 30\,000~K, when emissivities
of O and S ions are yet relatively weak.
This is also seen on Figure~\ref{FigLineFS}(b,c) above, where
the emissivities of the lines of \ion{Si}{iii} are a factor of 20-50
higher than that of \ion{O}{iii} and \ion{S}{iii} in the same conditions.

Resembling the case of oxygen recombination lines,
the density dependences are not pronounced.

\subsection{Ratios to collisionally-excited lines}

Comparison of metal recombination line fluxes with the ``traditional''
collisionally-excited lines may also be a useful tool for plasma diagnostics.

As these two types of lines have different origin, their 
emissivity dependences on temperature and density are
rather different.
For example, collisionally-excited line emissivities exponentially drop at
temperatures below about $h\nu/k$, as thermal electron energies are not
sufficient to excite the ion electronic transition.
In contrast, recombination line emissivities at low temperatures
increase with the temperature decreasing.

Also a useful property is that,
even in the case when the recombination lines are not detected at all,
from the limits on the ratios to the collisional lines it is possible
to put constraints on the plasma parameters.

\begin{figure}
\begin{center}
\centerline{
    \rotatebox{270}{
       \includegraphics[height=0.95\linewidth]{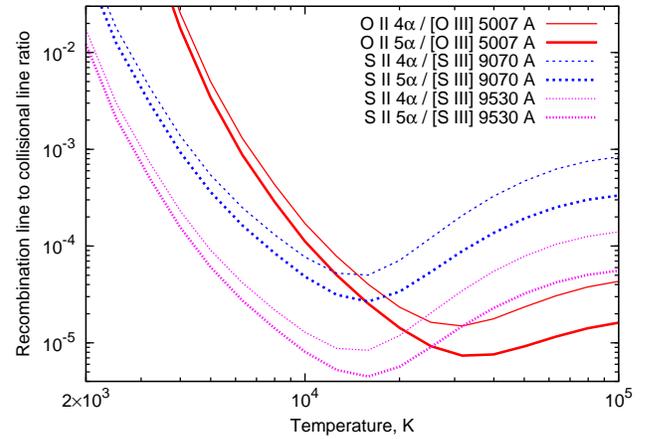}
                   }
           }
\caption{Low-density emissivity ratios of recombination lines
         to several brightest optical collisionally-excited lines.
         }
   \label{Figcrat}
  \end{center}
\end{figure}

On the Figure~\ref{Figcrat} we give ratios to several brightest
collisionally-excited optical lines.
The exponential increase of ratios at low temperatures is clearly seen.
Collisional line emissivities have been computed using Chianti
atomic database~\citep{Chianti,Chianti5}

\begin{figure}
\begin{center}
\centerline{
    \rotatebox{270}{
       \includegraphics[height=0.95\linewidth]{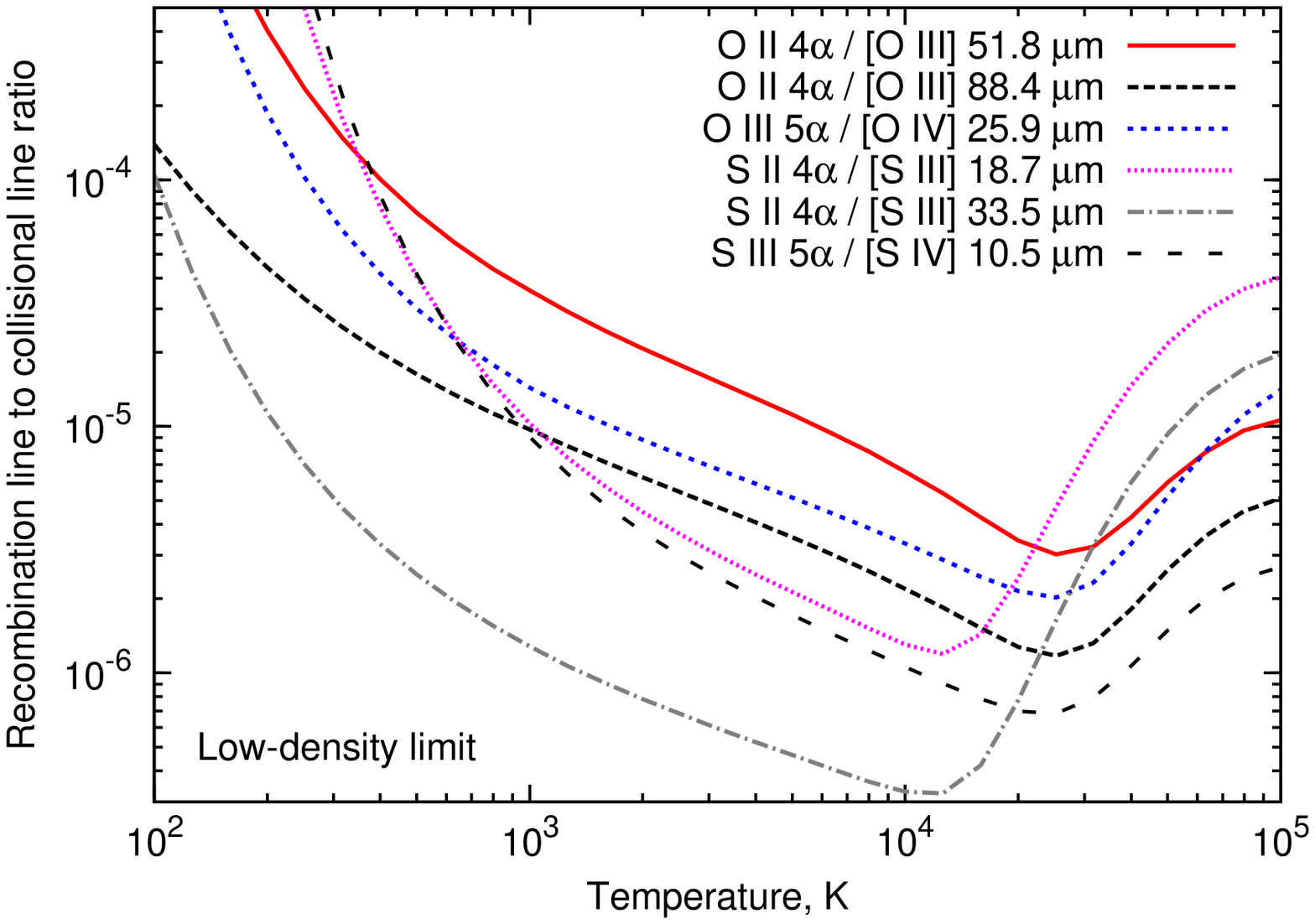}
                   }
           }
\centerline{
    \rotatebox{270}{
       \includegraphics[height=0.95\linewidth]{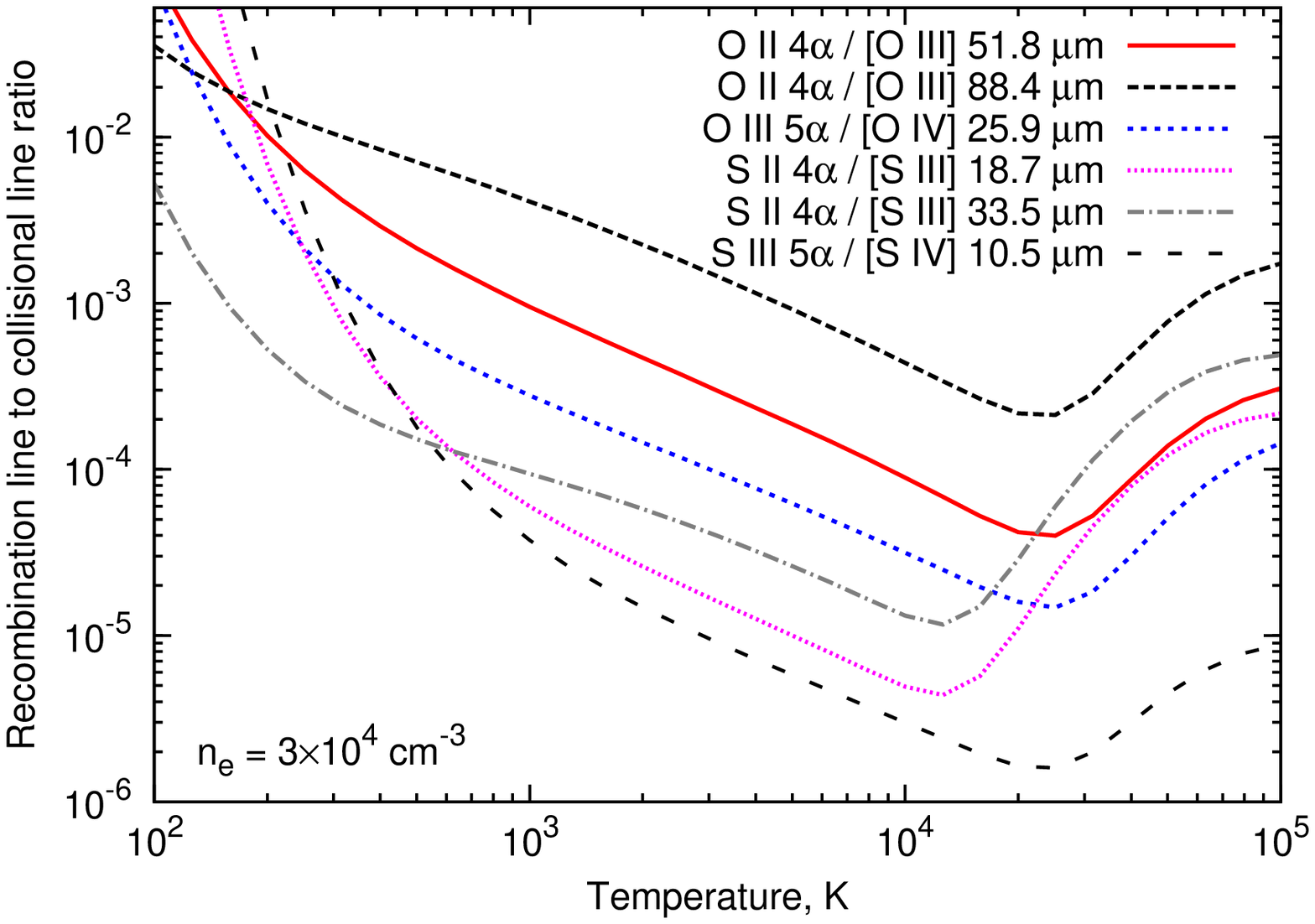}
                   }
           }
\caption{Emissivity ratios of recombination lines
         to several far-infrared collisionally-excited lines.
         Upper and lower panels show emissivity ratios
         in the low-density limit applicable for the cold photoionized region
         (except for the 88.4~$\mu$m line)
         and the case of $n_{\rm e}=3\times10^4$~cm$^{-3}$ typical
         for the cooling region after the shock front.
         }
   \label{Figcmrat}
  \end{center}
\end{figure}

Ratios to the fine-structure far-infrared line emissivities
are given on Figure~\ref{Figcmrat}.
As the Chianti database does not allow to compute line emissivities
down to 100~K, we computed them by extrapolating fine-structure transition
electronic excitation collision strengths to low temperatures by a constant,
that should be reliable to within a factor of two.
The collision strength values were adopted from calculations by
\citet{OIIIFS94,OIVFS92,SIIIFS99,SIVFS00,OIVFS06}.

We note that observations of the far-infrared lines
in the spectral range from 10 to 100~$\mu$m are impossible from the ground.
But even from space, observations of these very intense lines
cannot be performed with angular resolution sufficient
for resolving individual knots.
They can result only in signal integrated over many individual
emitting objects.

Density dependences of recombination and collisionally-excited
lines are also different.
Forbidden line emissivity starts decreasing as $1/n_{\rm e}$ at
some critical density,
whereas the recombination line emissivities may both decrease and increase
depending on the plasma temperature (see above).

The best density indicators for relatively low temperatures and densities
in the fast-moving knots are obviously ratios
of different optical and near-infrared line
to the far-infrared lines that have critical densities of the order of
$100-10^4$~cm$^{-3}$.
In the accompanying paper (Docenko \& Sunyaev, to be submitted),
we present such analysis based on existing experimental data
and compare the results with our predictions based on the SD95 model.

\section{Conclusions}
\label{SecConclusions}

We have offered and developed in detail a new method of
rapidly recombining plasma diagnostics based on measurements
of optical and near-infrared recombination lines of multiply-ionized
metal atoms.

As a promising example, we have applied our method to the SD95
theoretical model of fast-moving knots in Cassiopeia~A supernova remnant
and computed expected oxygen line fluxes from a single FMK of average size
and resulting recombination line ratios.
It turned out that both cold photoionized region before the shock front
and rapidly cooling region immediately after the shock front 
produce oxygen recombination lines strong enough to be observed on modern
optical telescopes in the wavelength range between 0.5 and 3~$\mu$m.

At shorter wavelengths, two reasons hamper the line observations:
 high absorption in the interstellar medium 
 on the way from Cas~A
and
 the line splitting into widely separated multiple components
 with consequently lower intensities in each of them.
The lines $n\alpha$ corresponding to transitions between
$n\ge6$ levels are the most promising,
as they are not so strongly split and 
have 50-90\% of intensity in one or a few narrow components.

The precision of our RL flux estimates from the Cas~A is not expected
to be better than a factor of several, as inconsistencies of similar magnitude
are observed between the SD95 model predictions and far-infrared
oxygen line observations (Docenko \& Sunyaev, to be submitted).

Nevertheless, when detected, 
the recombination lines will allow to
determine the details of the photoionization and rapid cooling processes
in the FMKs
from the line intensities, intensity ratios to each other and 
to forbidden collisionally-excited lines
and from the recombination line fine structure measurements.
The measurements of the line structure
demand higher signal-to-noise ratios
that can nevertheless be achieved on modern telescopes
in a reasonable integration time of few hours.

One very interesting result is that the predicted \ion{O}{v}
and \ion{O}{vi} recombination lines, being brightest
in the cooling region recombination line spectra,
arise in the temperature range significantly below $10^5$~K,
where there is essentially no O$^{5+}$ and O$^{6+}$ ions
in the collisional ionization equilibrium conditions.

It would be also very interesting to observe the recombination lines
from the knots consisting mostly of silicon and sulphur.
Existence of high-temperature shock-heated plasma having
such chemical composition have been proved by X-ray 
spectral observations (e.g., \citet{ChandraCasA2000,CasACha1Ms}),
but detailed theoretical predictions of cooling and ionization
structure of such clouds have not been developed yet.


{One of the purposes of this article is to attract attention
to the metal recombination lines radiated by plasma being far out of
the collisional ionization equilibrium and having highly-charged
ions at low temperatures. Besides the supernova remnants, such plasma
producing detectable metal recombination lines may exist in 
planetary nebulae, as well as near quasars and active galactic nuclei.}

\begin{acknowledgements}

We are grateful to L.A.~Vainshtein for 
  valuable advices,
  providing results from the ATOM computer code 
  and permission to use it.
We are also thankful to the anonymous referee for 
remarks permitting to make the paper easier to understand
and for a useful reference on computation of hydrogenic radiative
transition rates.
\end{acknowledgements}

\begin{appendix}

\section{Determination of the upper cutoff $n_{\rm max}$}
\label{AppA}

Traditionally, the highly-excited level populations are characterized
by the so-called departure coefficients $b_{nl}$, defined
as the ratio of actual level population $N_{nl}$ to its thermodynamic
equilibrium value $N_{nl}^*$.
With $n$ increasing, at levels $n> n_1$ the collisional 
$l$-redistribution processes establish equilibrium population over
$l$'s, i.e., $b_n=b_{nl}$.
The value of $n_1$ depends on the ion,
 as well as on electron temperature and number density.

\begin{figure}
\begin{center}
\centerline{
    \rotatebox{270}{
       \includegraphics[height=0.95\linewidth]{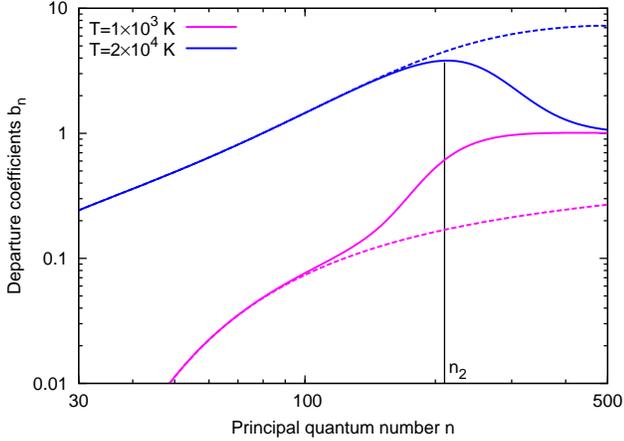}
                   }
           }
\caption{Departure coefficients $b_n$ of recombining
         \element[5+]{O} ions computed accounting for all processes
         (solid lines) and neglecting collisional $n$-redistribution,
         three-body recombination and collisional ionization
         (dashed lines)
         at electron temperatures 
         $T_{\rm e}=1\times10^3$~K and $2\times10^4$~K
         and density $n_{\rm e}=5\times10^4$~cm$^{-3}$.
         At higher temperature dielectronic recombination
         dominates and 
         significantly increases the level populations.
        }
   \label{FigOrec1}
  \end{center}
\end{figure}

For higher $n$'s, rates of collisional transitions involving $n$ change
start rapidly increasing and from some $n> n_2$ determine the
highly-excited 
level populations. Thus the $b_n$ curve,
itself defined only at $n>n_1$,
has two distinct ranges as a function of $n$:
below $n=n_2$ it is determined by recombination and radiative processes
and above it rapidly tends to unity because of the processes
relating different $n$ level populations with each other
and with the continuum: 
collisional $n$-redistribution,
three-body recombination and collisional ionization.
As an example relevant for our study, $b_n$'s of Li-like ion \element[5+]{O} 
at temperatures 1000 and 20\,000~K and electron density 50\,000~cm$^{-3}$
are shown on Figure~\ref{FigOrec1}, assuming $b_n=b_{nl}$.

On this figure, we also show the departure coefficients, computed
accounting only for radiative $n\to n'$ transitions.
It is seen that such approximation results in population computation
errors at levels higher than about $n_2$.

Let us now show that the electrons that recombined to levels
$n> n_2$ are contributing relatively weakly to the total recombination
and it is safe to introduce an upper cutoff $n_{\rm max}$,
neglecting population of higher levels.
Two cases -- $b_{n_2}>1$ and  $b_{n_2}<1$ -- should 
be discussed separately here.

\begin{figure}
\begin{center}
\centerline{
    \rotatebox{270}{
       \includegraphics[height=0.95\linewidth]{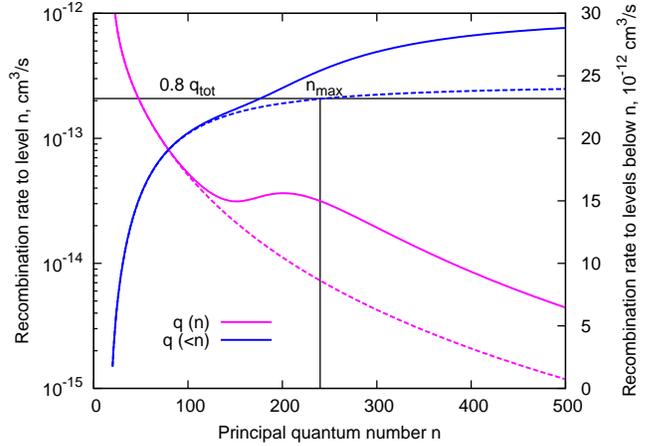}
                   }
           }
\centerline{
    \rotatebox{270}{
       \includegraphics[height=0.95\linewidth]{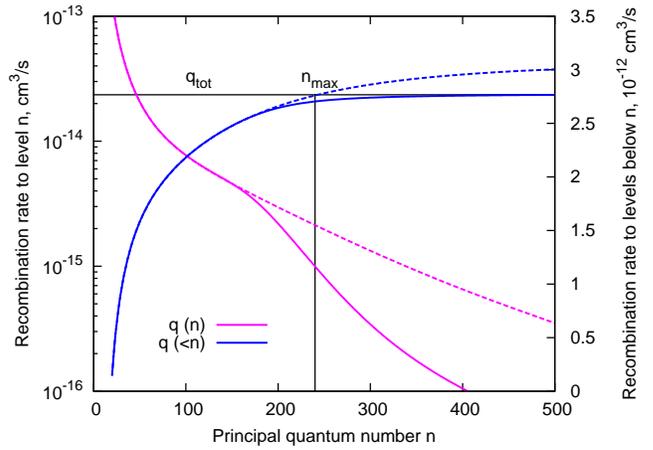}
                   }
           }
\caption{Recombination rates $q$ of \element[5+]{O}
         for $n_{\rm e}=5\times10^4$~cm$^{-3}$ and 
         $T_{\rm e}=1\times10^3$~K (upper panel) and 
         $2\times10^4$~K (lower panel).
         Meaning of solid and dashed lines is as on Figure~\ref{FigOrec1}.
         Plot shows both rates for a given $n$ (magenta curves)
         and cumulative rates for all levels below $n$ (blue curves). 
         Vertical line show upper cutoff $n_{\rm max}$ used for this ion
         in our calculations.
        }
   \label{FigOrec2}
  \end{center}
\end{figure}

In case of $b_{n_2}>1$ not introducing such cutoff 
and neglecting $n\to n'$ collisions may result in
some cases in significant overestimate of the recombination rate --
up to a factor of several.
The easiest estimate of the cutoff position
is the level where the total recombination rate onto $n< n_{\rm max}$
computed without accounting for $n\to n'$ transitions
equals to the true total recombination rate.
It is easy to understand that in this case $n_{\rm max}\approx n_2$,
as illustrated in the lower panel of Figure~\ref{FigOrec2}.

In case of $b_{n_2}<1$ the rates obtained neglecting collisional
transitions between high levels 
will always underestimate true recombination rates.
Though for temperatures above about $10^3$~K taking upper cutoff value 
approximately equal to $n_2$ does not result in error more than about 20\%
for relevant ions,
as illustrated on the upper panel of the Figure~\ref{FigOrec2}.
Lower cutoff values will result in more significant underestimates
of the total recombination rate and the line emissivities.

In case of lower densities, when $n_2$ is very high, another
natural limit $n_{\rm max}\approx (100-200)$ arises 
from general dependence of the DR rate on the principal quantum number,
see, e.g., \citet{Sobelman80}.

\section{Atomic physics for level population computations}
\label{AppB}

\subsection{{Radiative recombination}}
\label{SecRR}

There are two major recombination processes populating high-$n$
states in low-density plasmas: radiative and dielectronic
recombination (RR and DR, respectively).
{In the radiative recombination process, a free electron is directly
recombining to a bound state $(nl)$. A photon is simultaneously
emitted, carrying away the released energy.}

The radiative recombination level-specific cross sections 
$\sigma_{\rm RR}$
were computed from hydrogenic photoionization cross sections
$\sigma_{\rm PI}$ 
by applying the detailed balance relation:
\begin{equation}
  \sigma_{\rm RR} (\mathcal E\to nl) = 
    \frac{\alpha^2 Z^2}{2}\,
    \frac{2 l+1}{n^2}\,
    \frac{\nu^2}{\nu_n (\nu-\nu_n)}\,
  \sigma_{\rm PI}(nl\to \mathcal E),
\end{equation}
where $\mathcal E$ is the electron energy prior to recombination,
      $h\nu_n={\rm Ry}\, Z^2/n^2$ is the level binding energy,
      $h\nu=\mathcal E + h\nu_n$ is the energy of emitted photon,
      $Z$ is the recombining ion charge and 
      $\alpha$ is the fine-structure constant.
{The photoionization cross sections $\sigma_{\rm PI}$ were calculated
by the computer program of \citet{RADZ1}.}

{The level-specific radiative recombination rates
$q_{\rm RR}(nl;T_{\rm e}) \equiv 
   \langle \sigma_{\rm RR} (\mathcal E\to nl)\cdot\upsilon(\mathcal E)\rangle
$
were then computed by numerical integration over
the electron Maxwellian energy distribution for the temperature $T_{\rm e}$
(here $\upsilon$ denote electron velocity).}

{Using expressions from e.g.\ \citet{Spitzer56}, it may be shown
that the electron-electron collision timescale $\tau_{\rm eec}$
is always shorter than the cooling time $\tau_{\rm cool}$
at temperatures below $10^5$~K where the line emission occurs, 
and electrons have enough time to reach the Maxwellian distribution.
Roughly, 
$$
\tau_{\rm eec} = 2.66 \;{\rm s}\; \frac{T_4^{3/2}}
                                       {n_{\rm e,5} \ln \Lambda_{\rm C}} 
    \approx 0.16 \;{\rm s}\; \frac{T_4^{3/2}}{n_{\rm e,5}},
$$
where $T_{\rm e}=10^4T_4$~K, $n_{\rm e}=10^5n_{\rm e,5}$~cm$^{-3}$
and $\Lambda_{\rm C}\approx3.92\times10^7\; n_{\rm e,5}^{-1/2} T_4^{3/2}$.
}

\subsection{Dielectronic recombination}
\label{SecDR}

The DR is a two-step process 
(for reviews see, e.g., \citet{DRBVS,Shore69,DRSeaton,Sobelman80}),
illustrated in the simplest case by the following diagram:
\begin{equation}
  X^{Z+} (\gamma_0) + {\rm e} 
                     \rightleftarrows X^{(Z-1)+} (\gamma, nl) 
                     \to X^{(Z-1)+} (\gamma_0, nl) + \mbox{photon}. 
\end{equation}

Free electron e is first resonantly captured by the ion $X^{Z+}$ 
($Z$ denotes the ion charge)
with simultaneous core electron excitation 
from the ground state $\gamma_0$ to the state $\gamma$
having excitation energy $E_{\rm c}$.
This process is called dielectronic capture.
Resulting ion having two excited electrons is unstable with respect
to autoionization -- the inverse process of the dielectronic capture. 

The recombined ion then either autoionizes,
or its excited core electron transits back to the ground state $\gamma_0$
and emits a resonant line photon,
thus making the ion stable against the autoionization.

Therefore the level-specific dielectronic recombination rate
$q_{\rm DR}(\gamma, nl; T_{\rm e})$ is a product of two factors:
dielectronic capture rate and doubly-excited state stabilization probability.
In the simple case described above it is expressed as
\begin{equation}
\label{EqDR}
\begin{array}{l}
  q_{\rm DR}(\gamma, nl; T_{\rm e}) = 
    \left( \frac{2\pi\hbar^2}{mkT_{\rm e}} \right)^{3/2}
    \frac{g_\gamma (2l+1)}{g_0}
    \frac{A_{\rm c} A_{\rm a}(\gamma, nl; \gamma_0)}
         {A_{\rm c} + A_{\rm a}(\gamma,nl; \gamma_0)}
    \exp\left(-\frac{\mathcal E}{k T_{\rm e}}\right),
\end{array}
\end{equation}
where 
  $\gamma$ and $\gamma_0$ are recombining ion ground and excited states,
  $g_\gamma$ and $g_0$ are their statistical weights,
  $A_{\rm a}(\gamma,nl; \gamma_0)$ is the doubly-excited state $(\gamma,nl)$
    autoionization rate with core electron after the autoionization
    moving to the ground state $\gamma_0$,
  $A_{\rm c}$ is the core state $\gamma$ decay rate 
    to the ground state $\gamma_0$
  and $\mathcal E$ is the electron energy prior to the recombination,
  $\mathcal E = E_{\rm c} - {\rm Ry}\, Z^2/n^2$.

We have determined the dielectronic recombination rates using two methods,
which differ by the way they compute autoionization rates.
In the first method, the autoionization rate is expressed in terms of
the photoionization cross section using the dipole approximation
for the inter-electronic interaction.
Simple expression of this cross section and the autoionization rate
in the $n\gg 1$ limit
is then obtained using quasi-classical (QC) approach~\citep{PertAtom}.

It is known that the dipole approximation is not applicable 
for this purpose for $l\le3$ states~\citep{DRBVCh}. 
Though, this is not a serious limitation in the case, when
the dielectronic capture occurs via excitation
of the recombining ion core electron without change of its
principal quantum number ($\Delta n_{\rm c}=0$).
Such transitions are the most important in our case,
as relevant plasma temperatures are much lower than the
$\Delta n_{\rm c}>0$ transition excitation energies.

The second method is based on the usage of the program
ATOM~\citep{ShevelkoBook}.
It computes autoionization rates by extrapolating
below the threshold
the electronic excitation cross section of the recombining ion core
transition.
This extrapolation method is based on the correspondence principle and
is thus precise in the limit of $n\gg 1$.
Both in QC approximation and ATOM approach, the autoionization rates
decrease as $n^{-3}$.

\begin{figure}
\begin{center}
\centerline{
    \rotatebox{270}{
       \includegraphics[height=0.95\linewidth]{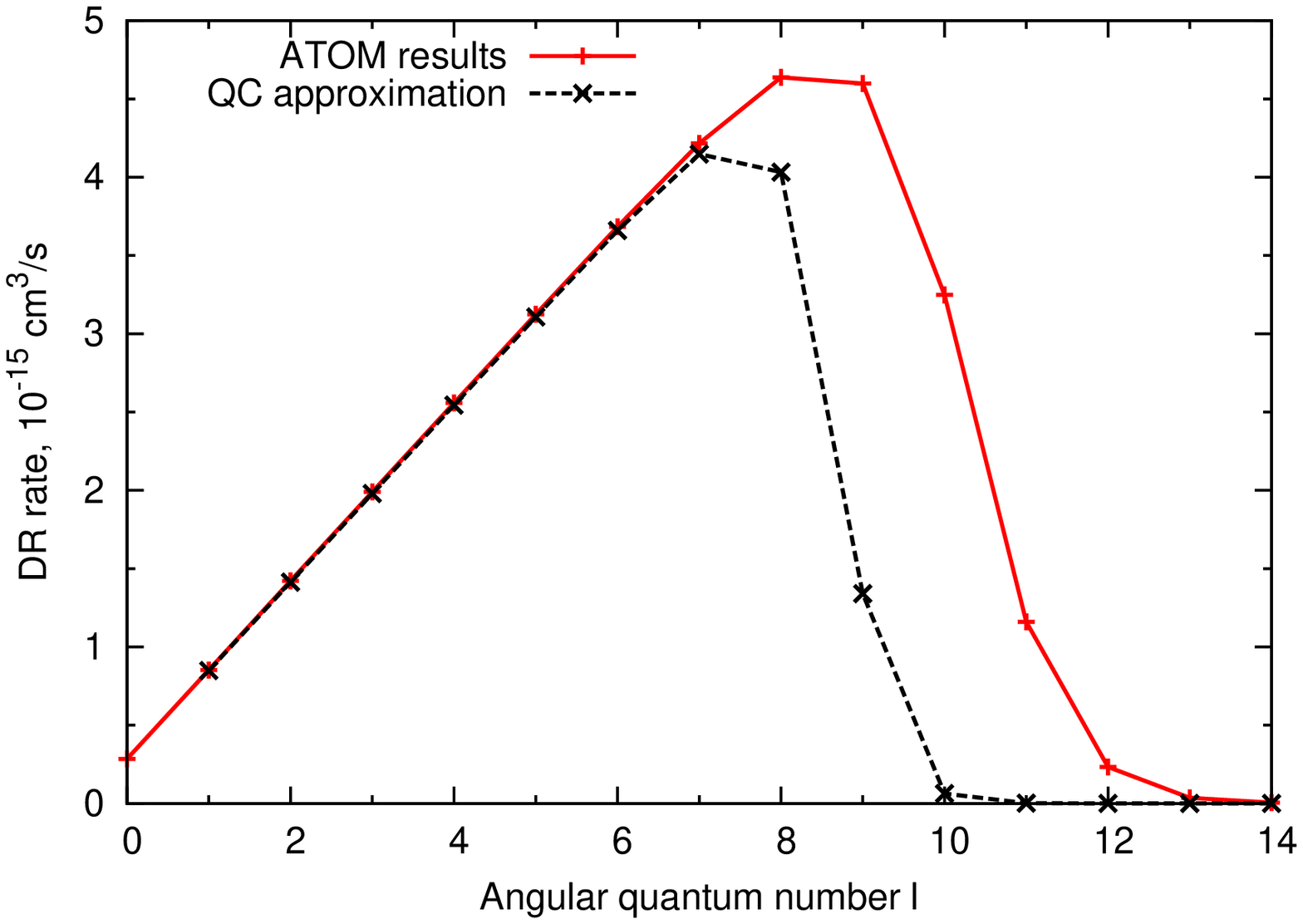}
                   }
           }
\centerline{
    \rotatebox{270}{
       \includegraphics[height=0.95\linewidth]{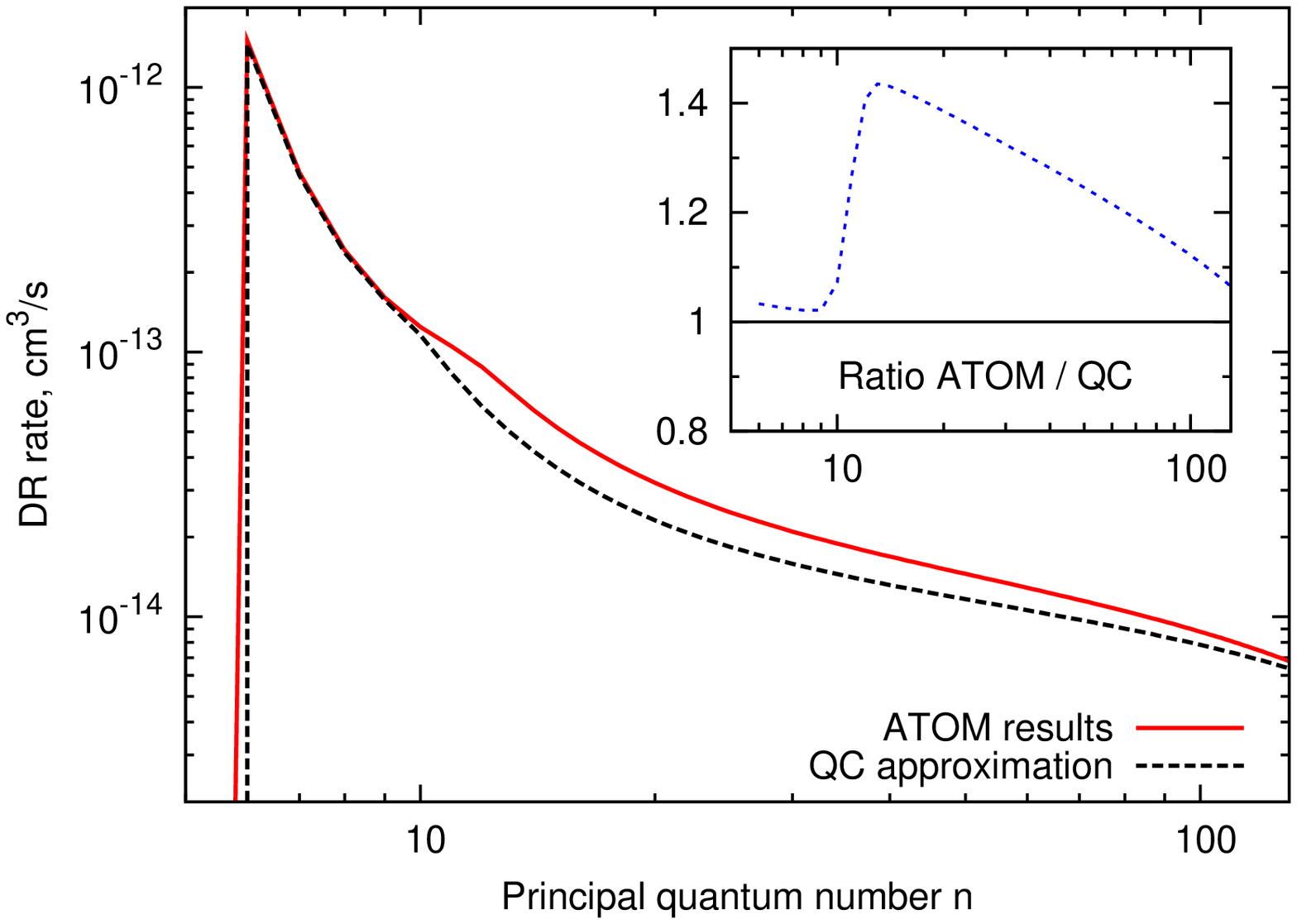}
                   }
           }
\caption{Comparison of $l$-resolved (upper panel) and 
         $l$-summed (lower panel) \element[5+]{O} 
         ion dielectronic recombination rates
         computed by the two methods described in the text:
         ATOM program and quasiclassical (QC) approximation. The values are
         given for $T_{\rm e}=2\times10^4$~K, but the curves scale identically
         with the temperature as long as $T_{\rm e}\ll10^6$~K, 
         when $\Delta n_c>0$ core excitations start playing a role.
         The $l$-resolved rates are given for $n=20$.}
   \label{FigDRRcomp}
  \end{center}
\end{figure}

\begin{figure}
\begin{center}
\centerline{
    \rotatebox{270}{
       \includegraphics[height=0.95\linewidth]{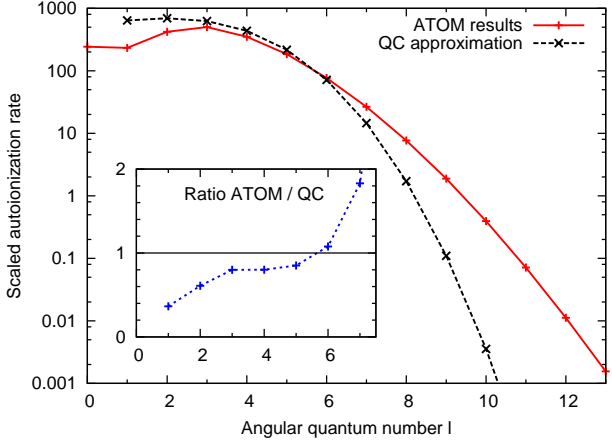}
                   }
           }
\caption{Comparison of scaled autoionization rates 
of \element[4+]{O} ion in $(2p,nl)$ state autoionizing to 
\element[5+]{O} ion in $2s$ state,
$A_{\rm a}(2p,nl;2s)\times n^3$, $10^{13}$ s$^{-1}$,
in the quasiclassical approximation and computed by the program ATOM.
        }
   \label{FigOVIWAI}
  \end{center}
\end{figure}

Comparison plots of the DR rates of \element[5+]{O} forming \element[4+]{O}
with recombined electron populating all $n$-levels
and $n=20$ level $l$-states
are given on Figure~\ref{FigDRRcomp}.
It is seen that in the quasiclassical approximation,
electrons are populating lower $l$'s, but higher $n$'s. 
As can be inferred from Eq.~(\ref{EqDR}), this corresponds
to sharper $l$-dependence of the autoionization rates.
Comparison of scaled quasiclassical and quantum autoionization rates
$A_{\rm a}(2p,nl;2s)\times n^3$ of doubly-excited oxygen ion \element[4+]{O}
is shown on Figure~\ref{FigOVIWAI},
indeed showing the inferred dependence.

It is clear from these data that the two models for the DR rates
will result in significantly different line emissivities 
(see also Figure~\ref{FigEmiss8a}).
For final results we decided to use the ATOM rates, where available,
as the theoretical arguments~\citep{ShevelkoBook}
show that they are more precise than the quasiclassical ones.

For ions
\element[5+]{O}, \element[4+]{O}, \element[4+]{Si}, \element[3+]{Si},
\element[2+]{Si}, \element[6+]{S}, \element[5+]{S} and \element[4+]{S}
we used the ATOM rates.
For other ions we implemented the quasiclassical expressions.

\begin{figure}
\begin{center}
\centerline{
    \rotatebox{270}{
        \includegraphics[height=0.95\linewidth]{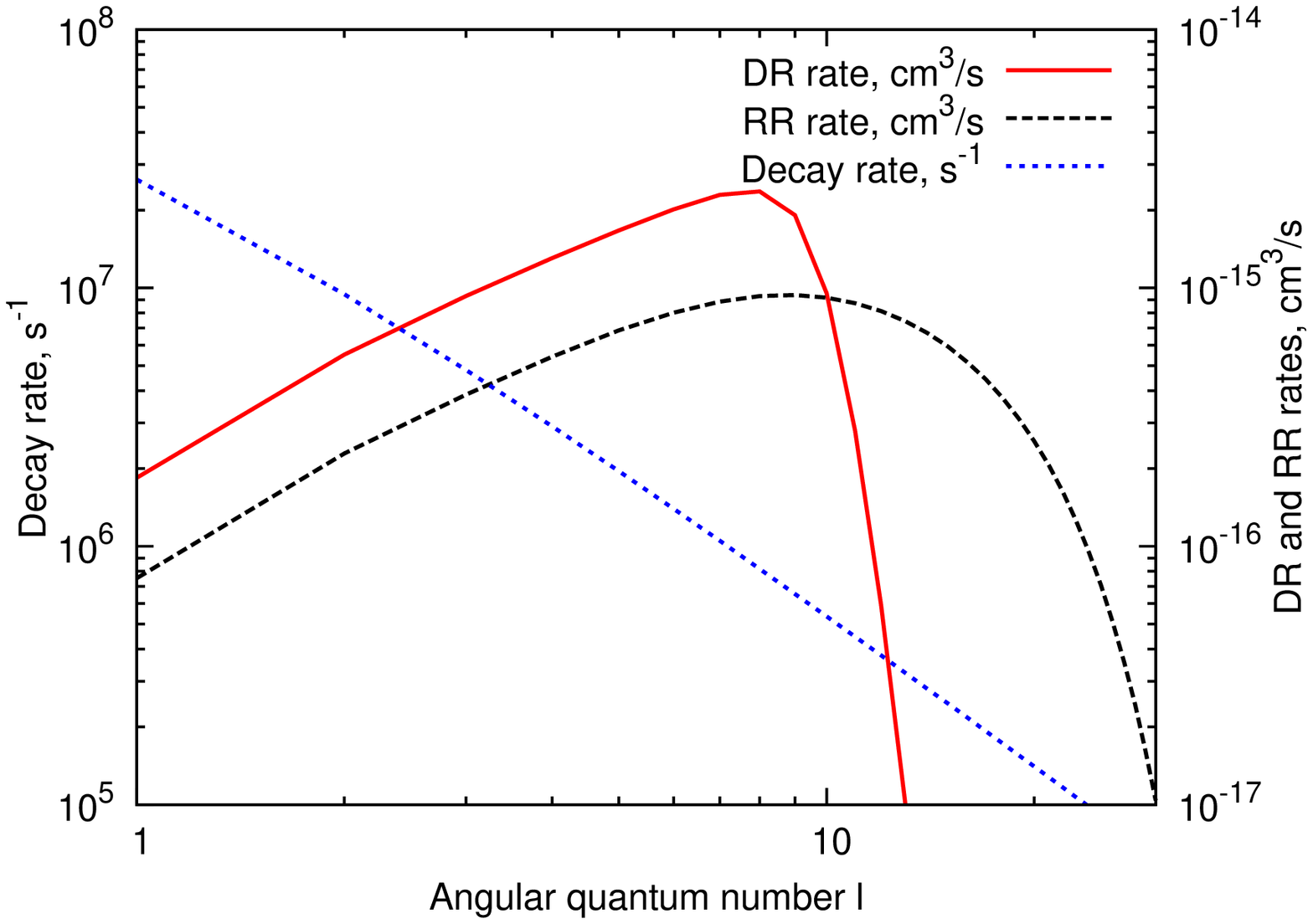}
                   }
           }
\centerline{
    \rotatebox{270}{
        \includegraphics[height=0.95\linewidth]{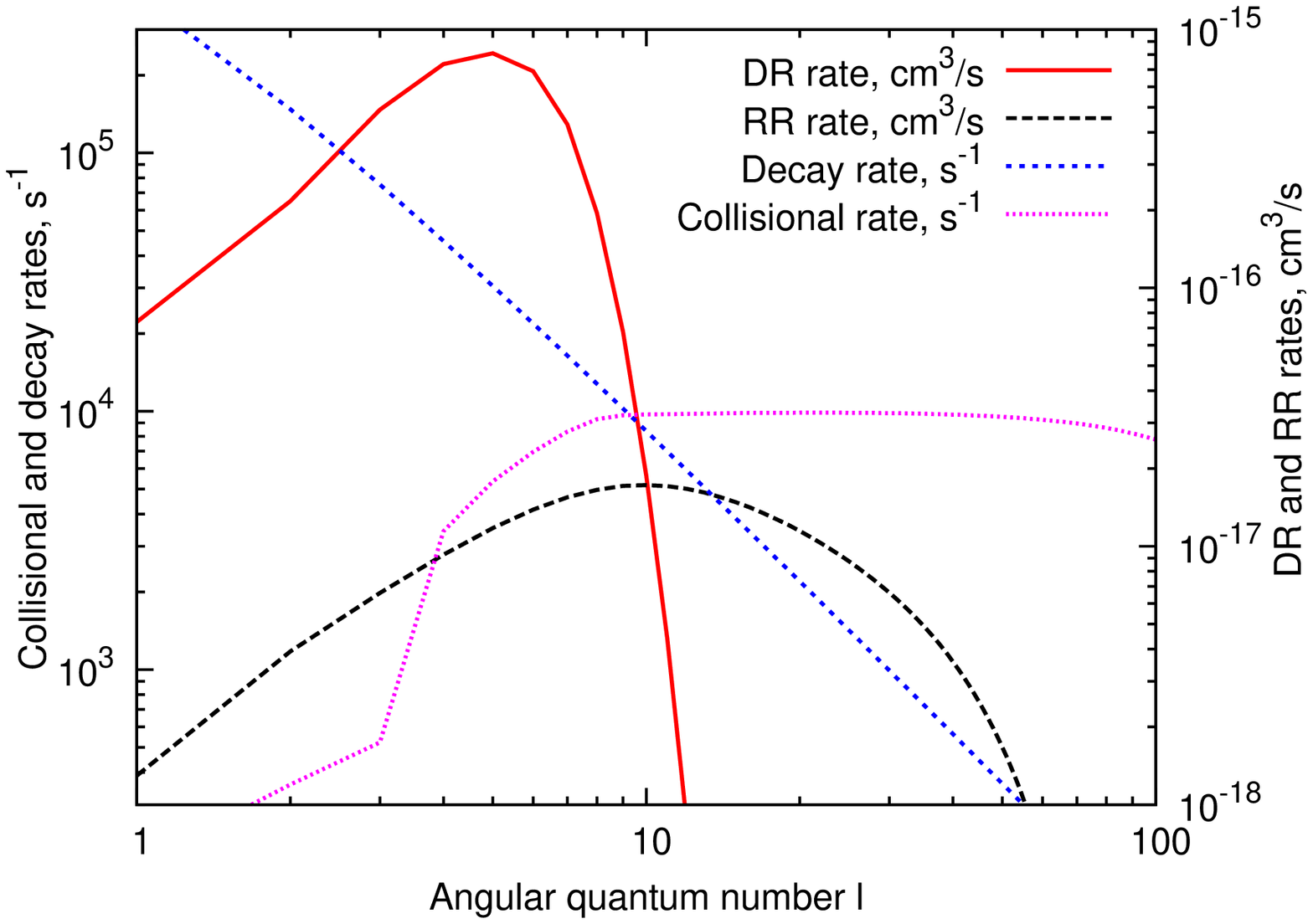}
                   }
           }
\caption{Elementary process rates for \element[5+]{O} ion recombining to
         form \element[4+]{O} with highly-excited 
         electron having 
         $n=50$ (upper panel) and $n=200$ (lower panel)
         at electron temperature $T_{\rm e}=2\times10^4$~K.
         Collisional rates (total for transitions to all $l'$) are
         shown for electron density $n_{\rm e}=5\times10^4$~cm$^{-3}$ and
         are negligibly low for $n=50$ case.
         }
   \label{FigEPOVI}
  \end{center}
\end{figure}

Comparison of the  dielectronic recombination rates
with other atomic process rates is shown on Figure~\ref{FigEPOVI}
for the case of \element[4+]{O} ion with highly-excited 
electron on levels $n=50$ and 200.
It is seen that the dielectronic recombination
populates lower $l$'s, but is more efficient at high-$n$ levels.
In case of sufficiently high density 
the dielectronic recombination to low-$l$ states
is followed by rapid redistribution onto much higher $l$'s
resulting in enhancement of the recombination line intensities.

\subsection{Highly-excited level energies}
\label{SecLevEn}

In hydrogen atom 
the energies of $(nl)$ levels are independent on $l$ because
of the shape of the Coulomb potential.
In atoms with several electrons the changes in the potential energy curve
introduce variations of the highly-excited level energies $E(nl)$ with $l$.
This dependence is usually parametrized by expression
$$
  E(nl) = {\rm Ry}\frac{z^2}{(n-\mu_{nl})^2},
$$
where $\mu_{nl}$ denotes quantum defect of the level $(nl)$
and $z=Z+1$ is the ion spectroscopic symbol.
The quantum defects $\mu_{nl}$ rapidly decrease with increasing $l$
and tend to constant with increasing $n$ \citep{SeatonQDT}.

Their values are important in this study for two reasons.
Firstly, level energies $E(nl)$ and $E(n'l')$ determine
transition energies $E(nl)-E(n'l')$ and, therefore,
spectral line wavelengths.
Transitions having the same values of $n$ and $n'$ have
similar wavelengths.
Small wavelength differences are in this case determined by
the energy differences within $n$ and $n'$ groups of states
-- by the quantum defects of the levels.

Secondly, the collisional transition rates $C_{nl,nl'}$ are 
dependent on the energy level splitting $|E(nl)-E(nl')|$,
growing as this splitting decreases.
Therefore the quantum defects in this case determine rates
of $l$-redistribution and recombination line flux
dependence on electron density in plasma.

In case of $l\le 2$ the quantum defects for most ions are known
and have been taken for our study from~\citet{NISTASD}%
\footnote{URL: {\tt http://physics.nist.gov/asd3}}.
In the next subsection we describe our method of computation
of level energy shift from hydrogenic values
$E_{\rm H}(nl)={\rm Ry}\, z^2/n^2$ for $l>2$
that can be expressed via quantum defects as
$$
 \Delta E(nl) \equiv  E(nl) - E_{\rm H}(nl)
               \approx 2{\rm Ry} \frac{z^2}{n^3}\mu_{l},
$$
where we have assumed that $\mu_{nl}=\mu_l$, i.e. that
the quantum defect does not depend on $n$ and that it is small:
$\mu_{l} \ll 1$.

The latter assumption is valid in our case of $l\gg 1$,
when the $(nl)$ states
are ``non-penetrating'', i.e., those in which highly-excited 
electron wave
function significantly differs from zero only outside the atomic core
region.
This property also allows to use hydrogenic expressions for
description of such highly-excited 
states.

\subsection{Computation of the level shifts for high-$l$ states}
\label{SecQuaDef}

Following the approach of~\citet{DRlikeC} and \citet{DickinsonC},
we account for the non-penetrating $(nl)$ state level shifts
arising due to two effects:
core polarizability
and electrostatic quadrupole interaction with the core.

The energy shift induced by the
core polarizability is proportional to the highly-excited 
electron radial integral
$\left\langle r_{nl}^{-4}\right\rangle$ and,
expressing it with hydrogenic formula, equals~\citep{SeatonQDT}
\begin{equation}
  \Delta E_{\rm p}(nl) = \frac{\alpha_{\rm c}}{a_0^3} \frac{z^4}{n^3}
    \frac{2(3-l(l+1)/n^2)}
         {l (l+1) (2l-1) (2l+1) (2l+3)}
    {\rm Ry}.
\end{equation}
Here $a_0$ is the Bohr radius and
     $\alpha_{\rm c}$ is the core polarizability, taken for relevant
     ions from the handbook of \citet{FragaHandbook} and review by
     \citet{Lundeen05} and given for reference in Table~\ref{TabForSub}.

\begin{table}
\caption{%
Atomic parameters used for computations of the line substructure.
\label{TabForSub}
}
\begin{center}
\begin{tabular}{lll|lll|lll}
\hline\hline
Ion & $\left\langle r_{\rm c}^2\right\rangle$ & $\alpha_{\rm c}$&
Ion & $\left\langle r_{\rm c}^2\right\rangle$ & $\alpha_{\rm c}$&
Ion & $\left\langle r_{\rm c}^2\right\rangle$ & $\alpha_{\rm c}$\\
\hline
\ion{O}{ii}  & 1.20 & 1.35 & \ion{Si}{ii}  & 4.33 & 11.7 & \ion{S}{ii} & 3.72 & 7.36 \\
\ion{O}{iii} & 0.97 & 0.94 & \ion{Si}{iii} & 3.79 & 7.42 & \ion{S}{iii}& 3.22 & 4.79 \\
\ion{O}{iv}  & 0.99 & 0.74 & \ion{Si}{iv}  &      & 0.20 & \ion{S}{iv} & 2.44 & 3.24 \\
\ion{O}{v}   & 0.89 & 0.27 &               &      &      & \ion{S}{v}  & 2.24 & 1.28 \\
\ion{O}{v}   &      & 0.007&               &      &      & \ion{S}{vi} &      & 0.07 \\
\hline
\end{tabular}
\end{center}
\emph{Note}.
Values are given in atomic units 
($a_0^2$ for $\left\langle r_{\rm c}^2\right\rangle$ 
and $a_0^3$ for $\alpha_{\rm c}$). Column ``ion'' contains the
ion spectroscopic symbol.
\end{table}

The electrostatic quadrupole interaction is absent
for core states having total electronic angular momentum $J_{\rm c}<1$
or total orbital momentum $L_{\rm c}<1$. 
If $J_{\rm c}\ge1$, the highly-excited 
electron interacts with it,
splitting the $(nl)$ level into $2J_{\rm c}+1$
components numbered by the quantum number $K$, where 
$\vec K = \vec J_{\rm c} + \vec l$.
Normally, if the $K$-splitting is present, it is larger than the
level shift due to core polarizability~\citep{Lundeen05}.

The quadrupole shift in this case is given by the
expression~\citep{Chang84}:
\begin{equation}
\label{EqdEq}
\begin{array}{l}
  \Delta E_{\rm q}(nl,K) = (-1)^{l+L_{\rm c}+S_{\rm c}+K} (2J_{\rm c}+1)
  \,{\left\{\begin{array}{ccc}
          J_{\rm c} & J_{\rm c} & 2 \\ 
          L_{\rm c} & L_{\rm c} & S_{\rm c} 
   \end{array} \right\}}
     \times \\
\quad \times
  \, {\left\{\begin{array}{ccc}
          J_{\rm c} & J_{\rm c} &  2\\ 
          l & l & K 
   \end{array} \right\}}
     \; (l||C^{(2)}||l) \; (L_{\rm c}||C^{(2)}||L_{\rm c})
     \; \left\langle r_{\rm c}^2 \right\rangle
     \; \left\langle r_{nl}^{-3} \right\rangle 2 {\rm Ry},
\end{array}
\end{equation}
where
 $S_{\rm c}$ is the total core spin momentum, 
 $\left\{:\;:\;:\right\}$ denote a $6j$-symbol,
 $\left\langle r_{\rm c}^2 \right\rangle$ and 
 $\left\langle r_{nl}^{-3} \right\rangle$
  are radial integrals in atomic units for core and highly-excited 
  electron, respectively,
  and  $(l||C^{(2)}||l)$'s are reduced matrix elements,
  given explicitly, e.g., by \citet{Sobelman1}:
$$
 (l||C^{(2)}||l) = - \sqrt{\frac{l(l+1)(2l+1)}{(2l+3)(2l-1)}}.
$$

The highly-excited 
electron radial integral can be determined using
hydrogenic expression from, e.g., \citet{BetheSalpeter}.
Core radial integrals $\left\langle r_{\rm c}^2 \right\rangle$
were computed for all relevant ions using radial wave functions
produced by the Flexible Atomic Code (FAC, \citet{FACref}).
Where literature data for the $\left\langle r_{\rm c}^2 \right\rangle$
were available~\citep{Sen79}, the differences between them and the FAC
values did not exceed about 10-15\%.
The used values of the $\left\langle r_{\rm c}^2 \right\rangle$
are given in Table~\ref{TabForSub}.

Expression~(\ref{EqdEq}) may be approximated in case
of P~core states ($L_{\rm c}=1$) and $l\gg 1$ by
\begin{eqnarray}
  \Delta E_{\rm q,\, max}(nl) & \approx & 0.2
     \, \left\langle r_{\rm c}^2 \right\rangle
     \, \left\langle r_{nl}^{-3} \right\rangle {\rm Ry}\nonumber \\
     & =& {} 0.2 \frac{z^3 \left\langle r_{\rm c}^2 \right\rangle}
                {n^3 l (l+1) (l+1/2)}
     {\rm Ry},
\end{eqnarray}
where in this case $\Delta E_{\rm q,\, max}(nl)$ gives maximum shift of all
the $K$-components. Components with both 
$\Delta E_{\rm q,\, max}(nl)$ and $-\Delta E_{\rm q,\, max}(nl)$
shifts are sometimes present.

Taking into account these effects, the total $(nl)$ state shift
for estimate of the collisional $l$-redistribution rates
was taken to be maximum possible, i.e., 
$$
\Delta E(nl) = \Delta E_{\rm q,\, max}(nl) + 
                \left|\Delta E_{\rm p}(nl)\right|.
$$
As larger shifts correspond to smaller collisional cross
sections, our approach may somewhat underestimate
the $l$-redistribution rates.

We estimate that the resulting uncertainties of level splittings,
in case of the $K$-splitting present,
correspond to the uncertainties in the electron density 
determination from the line ratios (see Section~\ref{SecRLdens})
being within a factor of about two.

\subsection{Computation of the line substructure}
\label{SecTLineStr}

As the ions of different elements have different values of $\alpha_{\rm c}$
and $\left\langle r_{\rm c}^2\right\rangle$,
the recombination line exact wavelengths will depend
not only on the ionic charge and the principal quantum numbers of the transition, 
but also on the element and the quantum numbers $l$, $l'$, $K$ and $K'$.

From the above expressions determining energy levels shifts it is possible
to calculate the positions of all line $nl\to n'l'$ components $nlK\to n'l'K'$.
Their relative fluxes are determined by the
so-called line strengths $S$ according to expression~\citep{Chang84}
\begin{equation}
 S(nlK,n'l'K') \propto (2K'+1) (2K+1) 
\,  {\left\{\begin{array}{ccc}
          K' & K & 1 \\ 
          l & l' & J_{\rm c} 
   \end{array} \right\}^2},
\end{equation}
if we assume that populations of the $(nlK)$-states are proportional to
their statistical weights equal to $(2K+1)$. 

It was shown by~\citet{Chang84} that $l=3$ levels in \ion{Si}{i}
are still partially penetrating the core. Therefore also
in our case the Si and S line substructure is described
adequately only for $l>3$.
In our paper we give results also for $l=3$,
but the precision of line position predictions in this case is
expected to be rather limited.

In the post-shock plasma of the fast moving knots,
the density is high enough to establish
significant population of lowest excited fine-structure
sublevels~\citep{FSpop79} that have typical excitation energies
corresponding to temperatures of several hundred Kelvin.

In some cases such excited state populations strongly change
the line substructure.
For example, in \element[3+]{O} ion, the ground state $^2P_{1/2}$ has
$J_{\rm c}=1/2$ and quadrupole interaction with it is absent.
Though, the lowest excited state $^2P_{3/2}$ of this ion has $J_{\rm c}=3/2$.
In case the core electron is in the excited state prior to recombination,
or getting there after the DR process, the highly-excited 
levels will have
$K$-splitting and resulting recombination lines will have much richer
substructure.

In our analysis we assume populations of the $J_{\rm c}$ sublevels
proportional to their statistical weights.

\subsection{Line emission without recombination}
\label{SecDCL}

We also account for the following process, mostly
not resulting in a recombination, but contributing, sometimes noticeably,
to the spectral line emissivities:
\begin{eqnarray}
  X^{Z+} (\gamma_0) + e & \rightleftarrows
      & X^{(Z-1)+} (\gamma, nl) \to \nonumber\\[3pt]
      &\to& X^{(Z-1)+} (\gamma, n'l') + {\rm photon}\to \\[3pt]
      &\to& X^{Z+} (\gamma_0) + e' + {\rm photon},\nonumber
\end{eqnarray}
i.e., highly-excited electron transition instead of core transition,
followed by the captured electron autoionization.
This process may be quite efficient
for increasing $\Delta n=1$ transition emissivities, as
the radiative transition rates for captured electron may be
significantly higher than that of the core electron,
if the core transition rate is low.

This is exactly the case for $\Delta n_{\rm c}=0$ transitions, mostly
contributing to the DR at relatively low temperatures.
Rate of the described process 
(``dielectronic capture with line emission'', DCL)
forming a line from transition $(nl)\to(n'l')$
may be easily expressed via the rate of the dielectronic recombination as
\begin{equation}
\label{EqDCL}
  q_{\rm DCL}(\gamma,nl,n'l'; T_{\rm e}) = q_{\rm DR}(\gamma,nl; T_{\rm e})\;
   \frac{A_{nl,n'l'}}{A_{\rm c}}.
\end{equation}

As the autoionization rates are much higher than the radiative transition
rates within the ion, probability of the DCL
is small and after the first captured 
electron transition it autoionizes, if it is energetically allowed.

To produce noticeable effect, highly-excited 
electron transition rates
should be at least comparable to the core transition rates, as
DCL does not produce cascades. 
Therefore, only transitions between lower $n$'s contribute
significantly to the additional emission in lines.
In case of \element[5+]{O}, described process increases the DR contribution
to the recombination line emissivity by roughly 5\% 
for the $8\alpha$ and $7\alpha$ lines.
From Eq.~(\ref{EqDCL}) 
it is clear that this fraction does not depend on temperature.

\subsection{Estimates of the resulting uncertainties}
\label{SecUncert}

Given the amount of approximations made to achieve the final results
-- the metal recombination line emissivities and wavelengths --
it might be useful to state explicitly resulting uncertainties
of these quantities.

The least reliable components for calculation of the recombination line
emissivities are dielectronic recombination rates and upper cutoff position.

For the ions where the ATOM results are available, the low-density 
emissivity and flux uncertainties are expected to be about 10\%.
For the ions where the calculations are made using the quasiclassical
expressions, the precision is lower and may constitute up to 50\%.

The choice of the upper cutoff $n_{\rm max}$ position results in additional
emissivity uncertainty of less than 10\% for the DR-dominated recombinations
and emissivity underestimate of less than 20\% for the RR-dominated case.

The density dependences of the line emissivities are not strong and should be
quite reliably modeled with additional uncertainties being below 5--10\%.

It should be reiterated that the absolute oxygen line fluxes are much less
confidently predicted, 
as their values are based on the theoretical FMK model
that shows differences when compared to the observed
optical-to-infrared collisional line ratios
up to a factor of several (Docenko \& Sunyaev, to be submitted).

The line fine structure uncertainties are coming from the 
input atomic data (polarizabilities and radial integrals) limited precision.
It is estimated to be not larger than 10--20\%, thus the line fine structure
should be quite reliable.

The recombination line fine structure component relative intensities
are the least reliably predicted as they depend
on the recombining ion ground state populations.
The results given above assume equilibrium populations;
this may not be true at low densities or temperatures.
Therefore the observed line structure may be significantly different
from one expected. From another side, observed line structure
will allow to put additional constraints on the electron density 
and temperature from the fine structure component intensities.

\end{appendix}


\bibliographystyle{aa}
\bibliography{Bibl_Doc}

\begin{thebibliography}{62}
\expandafter\ifx\csname natexlab\endcsname\relax\def\natexlab#1{#1}\fi

\bibitem[{{Baade} \& {Minkowski}(1954)}]{BaaMin54}
{Baade}, W. \& {Minkowski}, R. 1954, Astrophys.~J., 119, 206

\bibitem[{{Beigman} {et~al.}(1981){Beigman}, {Vainshtein}, \&
  {Chichkov}}]{DRBVCh}
{Beigman}, I.~L., {Vainshtein}, L.~A., \& {Chichkov}, B.~N. 1981, Soviet
  Physics JETP, 3, 490

\bibitem[{{Beigman} {et~al.}(1968){Beigman}, {Vainshtein}, \&
  {Sunyaev}}]{DRBVS}
{Beigman}, I.~L., {Vainshtein}, L.~A., \& {Sunyaev}, R.~A. 1968, Soviet Physics
  Uspekhi, 11, 411

\bibitem[{{Bethe} \& {Salpeter}(1957)}]{BetheSalpeter}
{Bethe}, H.~A. \& {Salpeter}, E.~E. 1957, {Quantum Mechanics of One- and
  Two-Electron Atoms} (New York: Academic Press)

\bibitem[{{Blum} \& {Pradhan}(1992)}]{OIVFS92}
{Blum}, R.~D. \& {Pradhan}, A.~K. 1992, Astrophys. J. Suppl. Ser., 80, 425

\bibitem[{{Borkowski} \& {Shull}(1990)}]{BorkowskiO}
{Borkowski}, K.~J. \& {Shull}, J.~M. 1990, Astrophys.~J., 348, 169

\bibitem[{{Brault} \& {Noyes}(1983)}]{Sun12mkm}
{Brault}, J. \& {Noyes}, R. 1983, Astrophys.~J., 269, L61

\bibitem[{{Bureyeva} \& {Lisitsa}(2000)}]{PertAtom}
{Bureyeva}, L.~A. \& {Lisitsa}, V.~S. 2000, Astrophysics and Space Physics
  Reviews, 11, 1

\bibitem[{{Chang}(1984)}]{Chang84}
{Chang}, E.~S. 1984, Journal of Physics B: Atomic and Molecular Physics, 17,
  L11

\bibitem[{{Chang} \& {Noyes}(1983)}]{ChangSun12mkm}
{Chang}, E.~S. \& {Noyes}, R.~W. 1983, Astrophys.~J., 275, L11

\bibitem[{{Chevalier} \& {Kirshner}(1978)}]{CheKirCasA}
{Chevalier}, R.~A. \& {Kirshner}, R.~P. 1978, Astrophys.~J., 219, 931

\bibitem[{{Chevalier} \& {Kirshner}(1979)}]{CheKirAbund}
{Chevalier}, R.~A. \& {Kirshner}, R.~P. 1979, Astrophys.~J., 233, 154

\bibitem[{{Dere} {et~al.}(1997){Dere}, {Landi}, {Mason}, {Monsignori Fossi}, \&
  {Young}}]{Chianti}
{Dere}, K.~P., {Landi}, E., {Mason}, H.~E., {Monsignori Fossi}, B.~C., \&
  {Young}, P.~R. 1997, Astron. Astrophys. Suppl. Ser., 125, 149

\bibitem[{{Dickinson}(1981)}]{DickinsonC}
{Dickinson}, A.~S. 1981, Astron. Astrophys., 100, 302

\bibitem[{{Dopita} {et~al.}(1984){Dopita}, {Binette}, \& {Tuohy}}]{Dopita84}
{Dopita}, M.~A., {Binette}, L., \& {Tuohy}, I.~R. 1984, Astrophys.~J., 282, 142

\bibitem[{{Fraga} {et~al.}(1976){Fraga}, {Karkowski}, \&
  {Saxena}}]{FragaHandbook}
{Fraga}, S., {Karkowski}, J., \& {Saxena}, K.~M.~S. 1976, {Handbook of Atomic
  Data} (Elsevier, Amsterdam)

\bibitem[{{Gerardy} \& {Fesen}(2001)}]{CasANIR01}
{Gerardy}, C.~L. \& {Fesen}, R.~A. 2001, Astron.~J., 121, 2781

\bibitem[{{Gordon}(1929)}]{Gordon29}
{Gordon}, W. 1929, Annalen der Physik, 394, 1031

\bibitem[{{Gu}(2003)}]{FACref}
{Gu}, M.~F. 2003, Astrophys.~J., 582, 1241

\bibitem[{{Heng} \& {McCray}(2007)}]{Kevin07}
{Heng}, K. \& {McCray}, R. 2007, Astrophys.~J., 654, 923

\bibitem[{{Hughes} {et~al.}(2000){Hughes}, {Rakowski}, {Burrows}, \&
  {Slane}}]{ChandraCasA2000}
{Hughes}, J.~P., {Rakowski}, C.~E., {Burrows}, D.~N., \& {Slane}, P.~O. 2000,
  Astrophys.~J., 528, L109

\bibitem[{{Hurford} \& {Fesen}(1996)}]{CasAFesen96}
{Hurford}, A.~P. \& {Fesen}, R.~A. 1996, Astrophys.~J., 469, 246

\bibitem[{{Hwang} {et~al.}(2004){Hwang}, {Laming}, {Badenes}, {Berendse},
  {Blondin}, {Cioffi}, {DeLaney}, {Dewey}, {Fesen}, {Flanagan}, {Fryer},
  {Ghavamian}, {Hughes}, {Morse}, {Plucinsky}, {Petre}, {Pohl}, {Rudnick},
  {Sankrit}, {Slane}, {Smith}, {Vink}, \& {Warren}}]{CasACha1Ms}
{Hwang}, U., {Laming}, J.~M., {Badenes}, C., {et~al.} 2004, Astrophys.~J., 615,
  L117

\bibitem[{{Inogamov} \& {Sunyaev}(2003)}]{KKHP-SK}
{Inogamov}, N.~A. \& {Sunyaev}, R.~A. 2003, Astronomy Letters, 29, 791

\bibitem[{{Itoh}(1981{\natexlab{a}})}]{Itoh81a}
{Itoh}, H. 1981{\natexlab{a}}, Publ. Astron. Soc. Japan, 33, 1

\bibitem[{{Itoh}(1981{\natexlab{b}})}]{Itoh81b}
{Itoh}, H. 1981{\natexlab{b}}, Publ. Astron. Soc. Japan, 33, 521

\bibitem[{{Itoh}(1986)}]{Itoh86}
{Itoh}, H. 1986, Publ. Astron. Soc. Japan, 38, 717

\bibitem[{{Klein} {et~al.}(2003){Klein}, {Budil}, {Perry}, \&
  {Bach}}]{CloudCrushExp03}
{Klein}, R.~I., {Budil}, K.~S., {Perry}, T.~S., \& {Bach}, D.~R. 2003,
  Astrophys.~J., 583, 245

\bibitem[{{Landi} {et~al.}(2006){Landi}, {Del Zanna}, {Young}, {Dere}, {Mason},
  \& {Landini}}]{Chianti5}
{Landi}, E., {Del Zanna}, G., {Young}, P.~R., {et~al.} 2006, Astrophys. J.
  Suppl. Ser., 162, 261

\bibitem[{{Lennon} \& {Burke}(1994)}]{OIIIFS94}
{Lennon}, D.~J. \& {Burke}, V.~M. 1994, Astron. Astrophys. Suppl. Ser., 103,
  273

\bibitem[{{Lundeen}(2005)}]{Lundeen05}
{Lundeen}, S.~R. 2005, Advances in Atomic, Molecular, and Optical Physics,
  Vol.~52, {Fine Structure in High-$l$ Rydberg states: A Path to Properties of
  Positive Ions} (Academic Press, edited by Chun C. Lin and Paul Berman),
  161--208

\bibitem[{{Malik} {et~al.}(1991){Malik}, {Malik}, \& {Varma}}]{MalikRnl}
{Malik}, G.~P., {Malik}, U., \& {Varma}, V.~S. 1991, Astrophys.~J., 371, 418

\bibitem[{{Mazzotta} {et~al.}(1998){Mazzotta}, {Mazzitelli}, {Colafrancesco},
  \& {Vittorio}}]{Mazzotta}
{Mazzotta}, P., {Mazzitelli}, G., {Colafrancesco}, S., \& {Vittorio}, N. 1998,
  Astron. Astrophys. Suppl. Ser., 133, 403

\bibitem[{{McKee} \& {Cowie}(1975)}]{McKeeCowie75}
{McKee}, C.~F. \& {Cowie}, L.~L. 1975, Astrophys.~J., 195, 715

\bibitem[{{Minkowski}(1957)}]{Mink57}
{Minkowski}, R. 1957, in IAU Symposium, Vol.~4, Radio astronomy, ed. H.~C. {van
  de Hulst}, 107

\bibitem[{{Minkowski} \& {Aller}(1954)}]{MinkAll54}
{Minkowski}, R. \& {Aller}, L.~H. 1954, Astrophys.~J., 119, 232

\bibitem[{{Peimbert} \& {van den Bergh}(1971)}]{Peimbert71}
{Peimbert}, M. \& {van den Bergh}, S. 1971, Astrophys.~J., 167, 223

\bibitem[{{Pengelly} \& {Seaton}(1964)}]{PengellySeaton}
{Pengelly}, R.~M. \& {Seaton}, M.~J. 1964, MNRAS, 127, 165

\bibitem[{{Pequignot} {et~al.}(1991){Pequignot}, {Petitjean}, \&
  {Boisson}}]{RRC91}
{Pequignot}, D., {Petitjean}, P., \& {Boisson}, C. 1991, Astron. Astrophys.,
  251, 680

\bibitem[{Ralchenko {et~al.}(2007)Ralchenko, Jou, Kelleher, Kramida, Musgrove,
  Reader, Wiese, \& Olsen}]{NISTASD}
Ralchenko, Y., Jou, F.-C., Kelleher, D., {et~al.} 2007, National Institute of
  Standards and Technology, Gaithersburg, MD

\bibitem[{{Schild}(1977)}]{RedLaw77}
{Schild}, R.~E. 1977, Astron.~J., 82, 337

\bibitem[{{Seaton}(1983)}]{SeatonQDT}
{Seaton}, M.~J. 1983, Reports of Progress in Physics, 46, 167

\bibitem[{{Seaton} \& {Storey}(1976)}]{DRSeaton}
{Seaton}, M.~J. \& {Storey}, P.~J. 1976, in Atomic processes and applications.
  P. G. Burke (ed.), North-Holland Publ. Co., Amsterdam, Netherlands, 133--197

\bibitem[{{Sen}(1979)}]{Sen79}
{Sen}, K.~D. 1979, Phys.~Rev.~A, 20, 2276

\bibitem[{{Shevelko} \& {Vainshtein}(1993)}]{ShevelkoBook}
{Shevelko}, V.~P. \& {Vainshtein}, L.~A. 1993, {Atomic physics for hot plasmas}
  (Bristol: IOP Publishing)

\bibitem[{{Shklovskii}(1968)}]{ShklovskyBook}
{Shklovskii}, I.~S. 1968, {Supernovae} (London, New York, etc.: Wiley)

\bibitem[{{Shore}(1969)}]{Shore69}
{Shore}, B.~W. 1969, Astrophys.~J., 158, 1205

\bibitem[{{Smeding} \& {Pottasch}(1979)}]{FSpop79}
{Smeding}, A.~G. \& {Pottasch}, S.~R. 1979, Astron. Astrophys. Suppl. Ser., 35,
  257

\bibitem[{{Sobelman}(1979)}]{Sobelman1}
{Sobelman}, I.~I. 1979, {Atomic spectra and radiative transitions} (Springer
  Series in Chemical Physics, Berlin: Springer)

\bibitem[{{Sobelman} {et~al.}(1981){Sobelman}, {Vainshtein}, \&
  {Yukov}}]{Sobelman80}
{Sobelman}, I.~I., {Vainshtein}, L.~A., \& {Yukov}, E.~A. 1981, {Excitation of
  atoms and broadening of spectral lines} (Springer Series in Chemical Physics
  7, Berlin: Springer)

\bibitem[{{Spitzer}(1956)}]{Spitzer56}
{Spitzer}, L. 1956, {Physics of Fully Ionized Gases} (New York: Interscience
  Publishers)

\bibitem[{{Storey} \& {Hummer}(1991)}]{RADZ1}
{Storey}, P.~J. \& {Hummer}, D.~G. 1991, Computer Physics Communications, 66,
  129

\bibitem[{{Summers}(1977)}]{Summersbnl}
{Summers}, H.~P. 1977, MNRAS, 178, 101

\bibitem[{{Sutherland} \& {Dopita}(1995{\natexlab{a}})}]{SD95I}
{Sutherland}, R.~S. \& {Dopita}, M.~A. 1995{\natexlab{a}}, Astrophys.~J., 439,
  365

\bibitem[{{Sutherland} \& {Dopita}(1995{\natexlab{b}})}]{SD95}
{Sutherland}, R.~S. \& {Dopita}, M.~A. 1995{\natexlab{b}}, Astrophys.~J., 439,
  381

\bibitem[{{Tayal}(2000)}]{SIVFS00}
{Tayal}, S.~S. 2000, Astrophys.~J., 530, 1091

\bibitem[{{Tayal}(2006)}]{OIVFS06}
{Tayal}, S.~S. 2006, Astrophys. J. Suppl. Ser., 166, 634

\bibitem[{{Tayal} \& {Gupta}(1999)}]{SIIIFS99}
{Tayal}, S.~S. \& {Gupta}, G.~P. 1999, Astrophys.~J., 526, 544

\bibitem[{{van den Bergh}(1971)}]{Bergh71}
{van den Bergh}, S. 1971, Astrophys.~J., 165, 457

\bibitem[{{Vrinceanu} \& {Flannery}(2001)}]{VrinStark}
{Vrinceanu}, D. \& {Flannery}, M.~R. 2001, Phys.~Rev.~A, 63, 032701

\bibitem[{{Watson} {et~al.}(1980){Watson}, {Western}, \&
  {Christensen}}]{DRlikeC}
{Watson}, W.~D., {Western}, L.~R., \& {Christensen}, R.~B. 1980, Astrophys.~J.,
  240, 956

\bibitem[{{Zel'dovich} \& {Raizer}(1967)}]{ZeldRaiz66}
{Zel'dovich}, Y.~B. \& {Raizer}, Y.~P. 1967, {Physics of shock waves and
  high-temperature hydrodynamic phenomena} (New York: Academic Press, edited by
  Hayes, W.D.; Probstein, Ronald F.)

\end{thebibliography}

\end{document}